\documentclass[12pt]{article}
\usepackage[utf8]{inputenc}
\usepackage{slashed}
\usepackage{epsfig}
\usepackage{comment}
\usepackage{feynmp}
\usepackage{cancel}
\usepackage{bbm}

\usepackage{amsfonts}
\usepackage{amsmath}
\usepackage{amssymb}
\usepackage{array}
\usepackage{bigints}
\usepackage{booktabs}
\usepackage[nosort]{cite}
\usepackage{color}
\usepackage{dsfont}
\usepackage{float}
\usepackage{framed}
\usepackage{graphicx}
\usepackage{indentfirst}
\usepackage{mathrsfs}
\usepackage{multirow}
\usepackage{setspace}
\usepackage{subdepth}
\usepackage{subfig}
\usepackage{titlesec}
\usepackage[dotinlabels]{titletoc}
\usepackage{wrapfig}
\usepackage[all]{xy}
\usepackage{young}
\usepackage[vcentermath]{youngtab}
\usepackage{relsize}
\usepackage{hyperref}
\hypersetup{colorlinks=true}
\hypersetup{linkcolor=black}
\hypersetup{citecolor=black}
\hypersetup{urlcolor=black}
\numberwithin{equation}{section}

\def\half{{1\over 2}}

\def\p{\partial}
\def\bz{{\bar z}}

\def\0{{(0)}}
\def\1{{(1)}}
\def\2{{(2)}}

\def\ci{{\mathscr I}}

\def\<{\langle }
\def\>{\rangle }

\def\p{\partial}

\newcommand{\bea}{\begin{eqnarray}}
\newcommand{\eea}{\end{eqnarray}}
\newcommand{\ba}{\begin{align}}
\newcommand{\ea}{\end{align}}

\renewcommand{\L}{\Lambda}

\newcommand{\w}{\omega}

\newcommand{\pd}[2]{\frac{\partial #1}{\partial #2}}

\newcommand{\bra}[1]{\left< #1 \right|}
\newcommand{\ket}[1]{\left| #1 \right>}

\newcommand{\beq}{\begin{equation}}
\newcommand{\eeq}{\end{equation}}
\newcommand{\beqa}{\begin{eqnarray}}
\newcommand{\eeqa}{\end{eqnarray}}
\newcommand{\beqar}{\begin{eqnarray*}}

\newcommand{\Or}{\mathcal{O}}

\newcommand{\ie}{{ i.e.}\ }

\newcommand{\ve}{{\varepsilon}}

\def\F{{\cal F}}
\def\A{{\cal A}}
\def\[{\[}
\def\]{\]}

\def\a{{\alpha}}
\def\da{{\dot \alpha}}
\def\db{{\dot \beta}}

\def\d{{\delta}}

\newcommand{\bd}[1]{\begin{fmffile}{#1}\begin{fmfgraph*}}
\newcommand{\ed}{\end{fmfgraph*}\end{fmffile}}
\renewcommand{\bar}{\overline}

\def\bar{\overline}
\def\b{\bar}

\def\half{{1 \over 2}}
\def\d{\partial}

\def\grad{\nabla}
\def\ep{\varepsilon}
\def\1{{\mathds 1}}

\def\alphadot{{\dot\alpha}}
\def\betadot{{\dot\beta}}

\newcommand{\Z}{{\mathbb Z}}
\newcommand{\C}{{\mathbb C}}
\newcommand{\R}{{\mathbb R}}

\def\SL{{\mathscr L}}

\def\CA{{\mathcal A}}
\def\CB{{\mathcal B}}
\def\CC{{\mathcal C}}
\def\CD{{\mathcal D}}

\def\CF{{\mathcal F}}

\def\CJ{{\mathcal J}}
\def\CK{{\mathcal K}}
\def\CL{{\mathcal L}}
\def\CM{{\mathcal M}}
\def\CN{{\mathcal N}}
\def\CO{{\mathcal O}}

\def\CQ{{\mathcal Q}}

\def\CS{{\mathcal S}}

\def\CV{{\mathcal V}}

\def\ie{\begin{equation}\begin{aligned}}
\def\fe{\end{aligned}\end{equation}}

\usepackage[left=2.5cm,right=2.5cm,top=2.5cm,bottom=3cm]{geometry}
\titleformat{\section}{\normalfont\bfseries}{\thesection.}{4pt}{}
\titlespacing{\section}{0pt}{20pt}{6pt}

\titleformat{\subsection}{\normalfont\itshape}{\thesubsection.}{4pt}{}
\titlespacing{\subsection}{0pt}{15pt}{6pt}

\titleformat{\subsubsection}{\normalfont}{\thesubsubsection.}{4pt}{}
\titlespacing{\subsubsection}{0pt}{15pt}{6pt}

\DeclareFontShape{OT1}{cmr}{mx}{n}%
    {<->cmr10}{}
\newcommand{\mytitlefont}{\fontseries{mx}\selectfont}
\DeclareMathAlphabet{\titlemath}{OT1}{cmr}{mx}{n}

\begin{document}


\begin{titlepage}

\begin{center}

~\\[2cm]

{\fontsize{25pt}{0pt} \mytitlefont Infinite-Dimensional Fermionic Symmetry}\\[0.1cm]
{\fontsize{26pt}{0pt} \mytitlefont

in Supersymmetric Gauge Theories}

~\\[0.5cm]

Thomas T.~Dumitrescu, Temple He, Prahar Mitra, and Andrew Strominger

~\\[0.1cm]

{\it Center for the Fundamental Laws of Nature,  \\
Harvard University, Cambridge, MA 02138, USA}

~\\[1.5cm]

\end{center}

\noindent We establish the existence of an infinite-dimensional fermionic symmetry in four-dimensional supersymmetric gauge theories by analyzing semiclassical photino dynamics in abelian~$\CN=~1$ theories with charged matter. The symmetry is parametrized by a spinor-valued function on an asymptotic~$S^2$ at null infinity. It is not manifest at the level of the Lagrangian, but acts non-trivially on physical states, and its Ward identity is the soft photino theorem. The infinite-dimensional fermionic symmetry resides in the same~$\CN=1$ supermultiplet as the physically non-trivial large gauge symmetries associated with the soft photon theorem. 
\vfill

\begin{flushleft}
November 2015
\end{flushleft}

\end{titlepage}


\pagestyle{empty}
\pagestyle{plain}

\tableofcontents
\bibliographystyle{utphys}

\section{Introduction}

Amplitudes that describe the emission or absorption of soft, massless particles by a hard scattering process often factorize into hard and soft contributions. In favorable cases, this factorization takes a simple, universal form, which does not depend on the detailed structure of the theory. Such factorization results are known as soft theorems (see for instance~\cite{Weinberg:1995mt} and references therein). A prototypical example is the soft photon theorem, which states that the leading infrared (IR) behavior of an~$(n+1)$-particle amplitude~$\CM_{n+1}$ involving a soft photon of momentum~$p^\text{s}$ takes the form
\begin{equation}\label{softhm}
\CM_{n+1} \rightarrow {\mathscr S} \CM_n \qquad \text{as} \qquad p^\text{s} \rightarrow 0~.
\end{equation} 
Here~$\CM_n$ only involves the~$n$ hard particles, and~${\mathscr S} $ is a soft factor that depends on the electric charges and momenta of the hard particles, as well as the polarization and the momentum of the soft photon. (The generalization to theories with magnetically charged particles was described in~\cite{Strominger:2015bla}.)  The precise form of~${\mathscr S}$  is essentially determined by Lorentz and gauge invariance and will be reviewed below. Soft theorems govern the leading and subleading IR behavior of amplitudes in abelian or nonabelian gauge theories and gravity; some old and recent analyses are in~\cite{Low:1954kd, GellMann:1954kc, Low:1958sn, Weinberg:1965nx, Burnett:1967km,Gross:1968in,Jackiw:1968zza,White:2011yy, Cachazo:2014fwa,Casali:2014xpa,Schwab:2014xua,Bern:2014oka,He:2014bga,Larkoski:2014hta,Cachazo:2014dia,Afkhami-Jeddi:2014fia,Adamo:2014yya,Geyer:2014lca,Schwab:2014fia,Bianchi:2014gla,Broedel:2014fsa,Bern:2014vva,White:2014qia,Zlotnikov:2014sva,Kalousios:2014uva,Du:2014eca,Liu:2014vva, Rao:2014zaa,Bonocore:2014wua,Luo:2014wea,Broedel:2014bza,Schwab:2014sla,Chen:2014xoa,Chen:2014cuc,Larkoski:2014bxa,Vera:2014tda,DiVecchia:2015oba,Cachazo:2015ksa,Lipstein:2015rxa,Adamo:2015fwa,Klose:2015xoa,Volovich:2015yoa,Bianchi:2015yta,Bork:2015fla,DiVecchia:2015bfa,Guerrieri:2015eea,Alston:2015gea,Chin:2015qza,DiVecchia:2015srk,Brandhuber:2015vhm}. In general, the soft factor~${\mathscr S}$ is replaced by an operator that acts on the external legs of the hard amplitude~$\CM_n$. Universal soft theorems are robust against quantum corrections, which are typically absent or restricted to low loop orders in perturbation theory; see for instance~\cite{Bern:2014oka,Bianchi:2014gla,Larkoski:2014bxa,Brandhuber:2015vhm} and references therein. In this paper, we will only discuss tree-level amplitudes.  

The universal soft behavior of gauge boson amplitudes was recently traced back to the existence of infinitely many symmetries that act on asymptotic scattering states at Minkowskian null infinity, i.e.~asymptotic symmetries, whose Ward identities are equivalent to the soft theorems~\cite{as,as1,He:2014laa,Kapec:2014opa,hmps,Lysov:2014csa,Campiglia:2014yka,Kapec:2014zla,Mohd:2014oja, Campiglia:2015yka,Kapec:2015vwa,He:2015zea,Campiglia:2015qka,Kapec:2015ena, Avery:2015gxa, Strominger:2015bla, Campiglia:2015kxa,Avery:2015rga}. Typically, these asymptotic symmetries can be viewed as large gauge transformations, which do not vanish at infinity and therefore act non-trivially on physical states.\footnote{~Here we follow the terminology of~\cite{as}: large gauge transformations are assumed to act non-trivially on physical states, because they do not vanish sufficiently rapidly at the boundary of spacetime. However, they may be topologically trivial, i.e.~deformable to the identity gauge transformation.} For instance,  the symmetries that give rise to the leading soft photon theorem~\cite{as,hmps,He:2015zea,Campiglia:2015qka,Kapec:2015ena,Strominger:2015bla,Avery:2015rga} are parametrized by a function~$\ep(z, \b z)$ on an asymptotic~$S^2$ (with coordinate $z \in \C$) inside the null boundary of Minkowski space. The corresponding charges~${\mathscr E}[\ep]$ are higher-harmonic generalizations of the electric charge,\footnote{~The charges~${\mathscr E}[\ep]$ were denoted by~$Q_\ep$ in~\cite{hmps}. Here we choose a different symbol to avoid a clash with the supersymmetry generators~$Q_\alpha$.} to which they reduce when~$\ep(z, \b z) =1$. Transformations with non-constant~$\ep(z, \b z)$ inhomogeneously shift the gauge field~$\CA_\mu$ by~$\d_\mu \ep$, and hence they are spontaneously broken. The corresponding Goldstone bosons are soft, zero-momentum photons.

It is natural to ask whether soft theorems for massless particles that are not gauge bosons have similar interpretations in terms of asymptotic symmetries. Here we will explore this question in the context of rigid supersymmetric gauge theories, where the gauge fields are accompanied by massless spin-$\half$ superpartners. For simplicity, we confine our attention to~$U(1)$ gauge theories with~$\CN=1$~supersymmetry and massless charged matter in four dimensions. The~$U(1)$ photon~$\CA_\mu$ has an electrically neutral, fermionic superpartner -- the photino~$\Lambda_\alpha$ -- whose couplings to charged matter are related to those of the photon by supersymmetry. The soft photino theorem takes the general form~\eqref{softhm}, where~${\mathscr S}$ is a non-trivial fermionic soft operator, which acts on the external states of the hard amplitude~$\CM_n$ in a universal fashion.

In this paper, we will establish the existence of infinitely many fermionic asymptotic symmetries, parametrized by a chiral spinor-valued function~$\chi_\alpha(z, \b z)$ on~$S^2$, whose Ward identities give rise to the soft photino theorem. The corresponding anticommuting charges~${\mathscr F}[\chi]$ act on the asymptotic fields at null infinity. However, unlike the infinity of bosonic charges~${\mathscr E}[\ep]$, they are not a subgroup of any obvious symmetry of the Lagrangian.\footnote{~The asymptotic symmetries related to the magnetic generalization of the leading soft photon theorem~\cite{Strominger:2015bla} or the subleading soft photon theorem~\cite{Lysov:2014csa} are also not manifest at the level of the Lagrangian.} The usual Lagrangian only displays a finite number of manifest fermionic symmetries -- the global supersymmetries generated by~$Q_\alpha$ and~$\b Q_\alphadot$. It is perhaps surprising that even rigid supersymmetric gauge theories can support an infinite number of fermionic asymptotic symmetries.\footnote{
A similar phenomenon occurs in three-dimensional, supersymmetric Chern-Simons theory in the presence of a suitably supersymmetric boundary, which supports a supersymmetric Kac-Moody current algebra.  (As we will see below, the asymptotic symmetries~${\mathscr E}[\ep]$ and~${\mathscr F}[\chi]$ also give rise to just such a current algebra.)  The bosonic Kac-Moody symmetries are conventional gauge transformations that do not vanish at the boundary. The Kac-Moody fermions can be understood as a remnant of the full super gauge symmetry that is present before fixing Wess-Zumino (WZ) gauge (see for instance~\cite{Sakai:1989nh}). It is plausible that our asymptotic symmetries~${\mathscr F}[\chi]$ have a similar interpretation, but we will not show it here. Instead, we will exhibit the charges~${\mathscr F}[\chi]$ directly in WZ gauge and explore their properties.} By contrast, this is expected in supergravity, where local supersymmetry is a gauge symmetry~\cite{ly,avsch}. 

Under the action of~${\mathscr F}[\chi]$ we find that the photino~$\Lambda_\alpha$ shifts inhomogenously. Hence these symmetries are spontaneously broken, and the soft photini are interpreted as the corresponding Goldstone fermions. Interestingly, supersymmetry relates the fermionic charges~${\mathscr F}[\chi]$ to the bosonic charges~${\mathscr E}[\ep]$. We find (see~\eqref{smii} below),
\begin{equation} \label{sm} 
\big\{\zeta^\alpha Q_\alpha, {\mathscr F}[\chi]\big\} = i {\mathscr E}\left[\zeta^\alpha \chi_\alpha\right]~, \qquad \big\{\b Q_\alphadot, {\mathscr F}[\chi]\big\} = 0~.
\end{equation}
Here the supersymmetry transformation is parametrized by a commuting, constant spinor~$\zeta_\alpha$, and~$\chi_\alpha(z, \b z)$ is also taken to be commuting. The charges~${\mathscr E}[\ep]$ commute with~$Q_\alpha$ and~$\b Q_\alphadot$.

The soft photon theorem implies that the insertion of a zero-momentum, positive-helicity photon into a scattering amplitude can be interpreted as the Ward identity for a~$U(1)$ Kac-Moody current, which transforms in a~$(1, 0)$ representation of the~$SL(2, \C)$ conformal symmetry acting on the~$S^2$ at null infinity~\cite{as,hmps}. Similarly, we will see that the insertion of a positive-helicity photino behaves like a~$(\half, 0)$ current on~$S^2$. The two currents are related by supersymmetry, as was the case for the charges in~\eqref{sm}.

In section~\ref{sec:softhms}, we begin by reviewing basic aspects of abelian gauge theories with~$\CN=1$ supersymmetry, focusing on the structure of the supermultiplet that contains the electric current~$\CJ_\mu$, which couples to the photon, and its fermionic superpartner~$\CK^F_\alpha$, which couples to the photino. We present a current-algebra derivation of the tree-level soft photon and photino theorems that utilizes the properties of~$\CJ_\mu$ and~$\CK_\alpha^F$ matrix elements between asymptotic states. This derivation emphasizes the universality of the two soft theorems, as well as their relation via supersymmetry. It also clarifies the structure of the soft operator~${\mathscr S}$ in~\eqref{softhm} that arises in the soft photino theorem. 
 
In section~\ref{sec:asymni} we analyze the classical dynamics of the supersymmetric gauge theory near null infinity. This is facilitated by a convenient choice of coordinates and spinor basis, in which the asymptotic behavior of massless fields near null infinity is simply related to their quantum numbers with respect to the conformal group that governs the deep IR behavior of the theory. After reviewing the results of~\cite{hmps} on the asymptotic dynamics of the photon and the associated bosonic charges~${\mathscr E}[\ep]$, we repeat the analysis for the photino. We construct the fermionic asymptotic charges~${\mathscr F}[\chi]$ and establish some of their basic properties. 

In section~\ref{sec:sphotisjas} we show that the Ward identity for the fermionic symmetries~${\mathscr F}[\chi]$ reproduces the soft photino theorem derived in section~\ref{sec:softhms}.

Our conventions, as well as various useful formulas, are summarized in appendix~\ref{app:conventions}. Appendix~\ref{modeexp} describes the mode expansions of all fields that appear in this paper in terms of creation and annihilation operators.

\section{Soft Theorems}
\label{sec:softhms} 

\subsection{Aspects of~$\CN=1$ Gauge Theories}

Unless stated otherwise, we will use the conventions of~\cite{Wess:1992cp}. As was stated in the introduction, we will consider~$U(1)$ gauge theories with~$\CN=1$ supersymmetry.
In the $\CN=1$ superspace formalism, one starts by introducing a vector superfield which has scalar fields $\CB$ (complex), $\CC$, $\CD$ a vector field $\CA_\mu$ and fermions $\Psi_\a$, $\L_\a$ as its components. The usual $U(1)$ gauge transformations elevate to supergauge transformations parameterized by two complex scalars $F$, $\ve$ and a fermion $\psi_\a$. This acts on the component fields as
\begin{equation}
\begin{split}\label{supergaugetransform}
 ( \CB , \CC , \CD , A_\mu , \Psi_\a , \L_\a  )  \to  (  \CB + F , \CC - \text{Im} \,\ve , \CD , \CA_\mu + \p_\mu \text{Re} \, \ve , \Psi_\a +  \psi_\a ,  \L_\a  ) .
\end{split}
\end{equation}
It is clear that we can choose $F$, $\text{Im}\,\ve$ and $\psi_\a$ in such a way to set $\CB = \CC = \Psi_\a = 0$. This defines the Wess-Zumino (WZ) gauge. After fixing this gauge, the vector multiplet $\CV$ is given by 
\begin{equation}
\label{gaugemult}
\CV = \left(\CA_\mu\,, \Lambda_\alpha\,, \b \Lambda_\alphadot\,, \CD\right)~.
\end{equation}
Here~$\CA_\mu$ is the~$U(1)$ gauge field (i.e.~the photon) with field strength~$\CF_{\mu\nu} = \d_\mu \CA_\nu - \d_\nu \CA_\mu$. It is subject to conventional~$U(1)$ gauge transformations, which remain unfixed in WZ gauge. The spin-$\half$ superpartner of the photon is the photino, which is described by a left-handed Weyl fermion~$\Lambda_\alpha$ and its right-handed Hermitian conjugate~$\b \Lambda_\alphadot$. The vector multiplet also contains a real scalar~$\CD$, which is a non-propagating auxiliary field. 

At this stage, we pause and reflect on the validity of choosing the WZ gauge. We must be careful to distinguish between small gauge transformations whose gauge parameters vanish on the boundary of spacetime from large gauge transformation which do not. Small gauge transformations represent redundancies in the theory and only they can be used to fix a particular gauge choice. On the other hand, large gauge transformations generate asymptotic symmetries which are physical symmetries with non-trivial consequences for the scattering matrix. It is therefore important to check that the WZ gauge condition can be imposed via a small gauge transformation. There are two ways to do this. The first is to perform a thorough analysis of the boundary conditions of all the fields pre-gauge fixing. This can be done by working out the equations of motion for all the fields in the vector multiplet (using the superspace formalism). The boundary conditions of the fields are then determined by analysing these equations near $\ci$. Using this, one can check if small gauge transformations can be used to set $\CB = \CC = \Psi_\a = 0$.

The second approach (and the one we employ in this paper) is to work in WZ gauge and search for hints of large gauge symmetries through soft theorems. To understand why this works, we recall the well-established connection between the leading soft-photon theorem and $U(1)$ large gauge transformations \cite{hmps}. It is a standard assumption in quantum field theory that all fields -- and in particular, the gauge field $\CA_\mu$ -- vanish on the boundary of spacetime. As we now understand, such a boundary condition is in fact too restrictive and obscures the existence of an infinite-dimensional asymptotic symmetry group (namely large $U(1)$ gauge transformations). However, the existence of this symmetry is already hinted at due to the existence of soft theorems (and infrared divergences \cite{Kapec:2017tkm}) within standard quantum field theory. The soft theorems are Ward identities of these asymptotic symmetries and thereby contain information about them. Analogously, in our case by working in WZ gauge, it is entirely possible that we have obscured some large gauge symmetry in the theory but if this is the case, we will be able to see this via a soft theorem -- the \emph{only} candidate for which is the soft photino theorem. Indeed as mentioned in the introduction, a central goal of this paper is to use the soft photino theorem to establish the existence of infinitely many fermionic asymptotic symmetries. However, as we from \eqref{fsoftcomm}, the large gauge symmetries responsible for the soft photino theorem shift the photino, $\L_\a \to \L_\a + \sigma^\mu_{\a\da} \p_\mu {\overline \chi}^\da$. On the other hand, the supergauge transformations \eqref{supergaugetransform} leave the photino invariant. It follows that the gauge transformation required to impose the WZ gauge condition is \emph{not} a large gauge transformation!

In WZ gauge, the non-vanishing (anti-) commutators of the component fields in the vector multiplet~$\CV$ with the supercharges~$Q_\alpha, \b Q_\alphadot$ are given by 
\begin{subequations}\label{gaugesusy}
\begin{align}
 \big[Q_\alpha,  \CA_\mu \big] & = -  \sigma_{\mu\alpha\alphadot} \b \Lambda^\alphadot~,  &  \big[\b Q_\alphadot,  \CA_\mu \big] & =  \Lambda^\alpha \sigma_{\mu\alpha\alphadot}~,\\
 \big\{Q_\alpha,  \Lambda_\beta\big\} &= \ep_{\alpha\beta} \CD - i  \left(\sigma^{\mu\nu}\right)_{\alpha \beta}\CF_{\mu\nu}~,  & \big\{\b Q_\alphadot,  \b \Lambda_\betadot\big\} & = \ep_{\alphadot\betadot} \CD - i  \left(\b \sigma^{\mu\nu}\right)_{\alphadot \betadot}\CF_{\mu\nu}~,\\
 \big[Q_\alpha, \CD\big] & = - i \sigma^\mu_{\alpha\alphadot} \d_\mu \b \Lambda^\alphadot~,  &  \big[\b Q_\alphadot, \CD\big] & = -i \d_\mu \Lambda^\alpha \sigma^\mu_{\alpha\alphadot}~.
\end{align}
\end{subequations}
The dynamics of the gauge multiplet is described by a Lagrangian~$\SL_\text{gauge}$, which is invariant (up to a total derivative) under the supersymmetry transformations in~\eqref{gaugesusy},
\begin{equation}\label{action}
\SL_\text{gauge} = - \frac{1}{4e^{2}}  \F_{\mu\nu} \F^{\mu\nu} - \frac{i}{e^{2}} {\b \Lambda} \, \b \sigma^{\mu} \p_{\mu } \Lambda + \frac{1}{2e^{2}} {\cal D}^2 + \left(\text{higher-derivative terms}\right)~.
\end{equation} 
In addition to the standard two-derivative kinetic terms for the gauge multiplet, we are allowing for the possibility of higher-derivative terms, e.g.~terms such as~$F^4 + \left(\text{fermions}\right)$, which arise in supersymmetric Born-Infeld actions. The soft theorems discussed below remain valid in the presence of such terms.

The interaction of the gauge field~$\CA_\mu$ with matter proceeds through a conserved current~$\CJ_\mu$, which resides in a real linear multiplet~$\CJ$,\footnote{~In superspace, a real linear multiplet is described by a real superfield~$\CJ$ that satisfies the constraints~$D^2 \CJ = \b D^2 \CJ = 0$, where~$D_\alpha, \b D_\alphadot$ are the usual super-covariant derivative operators defined in~\cite{Wess:1992cp}.}
\begin{equation}\label{jmult}
\CJ = \left(\CK^B\,, \CK^F_\alpha\,, \b \CK^F_\alphadot\,, \CJ_\mu\right)~, \qquad \d^\mu \CJ_\mu = 0~.
\end{equation}
Here~$\CK^B$ is a real scalar, while~$\CK^F_\alpha$ and its Hermitian conjugate~$\b \CK^F_\alphadot$ are left- and right-handed Weyl spinors. Unlike~$\CJ_\mu$, neither~$\CK^B$ nor~$\CK^F_\alpha, \b \CK^F_\alphadot$ obey a differential constraint, i.e.~they are not conserved currents. All fields in the current supermultiplet~$\CJ$ are gauge invariant. Their non-vanishing supersymmetry transformations take the following model-independent form,
\begin{subequations}\label{currsusyvar}
\begin{align}
\big[Q_\alpha, \CK^B\big] & =  i \CK^F_\alpha~, & \big[\b Q_\alphadot, \CK^B\big] & =  i \b \CK^F_\alphadot~, \\
\big\{\b Q_\alphadot, \CK_\alpha^F\big\}& = - i \sigma^\mu_{\alpha \alphadot} \left(\CJ_\mu + i \d_\mu \CK^B\right)~, & \big\{Q_\alpha, \b \CK_\alphadot^F\big\}& = i \sigma^\mu_{\alpha \alphadot} \left(\CJ_\mu - i \d_\mu \CK^B\right)~,\\
\big[Q_\alpha, \CJ_\mu\big]  & = -2  {\left( \sigma_{\mu\nu} \right)_\alpha}^\beta \d^\nu \CK^F_\beta~, & \big[\b Q_\alphadot, \CJ_\mu\big]  & = 2  \left( \b \sigma_{\mu\nu} \right)_{\alphadot\betadot} \d^\nu \b \CK^{F\betadot}~.
\end{align}
\end{subequations}
At first order, the interaction of the fields in the vector multiplet~$\CV$ with matter proceeds via the following universal couplings to the operators in the current multiplet~$\CJ$,\footnote{~For instance, this means that~$\displaystyle \CJ_\mu(x)= -{\delta S_\text{int} \over \delta \CA^\mu(x)}$, where~$S_\text{int}  = \int d^4 x \, {\mathscr L}_\text{int}$.}
\begin{equation}\label{matintlag}
\begin{split}
\SL_{\text{int}}  = - \CA^\mu \CJ_\mu - i \Lambda \CK^F + i \b \Lambda \, \b \CK^F - \CD \CK^B + \left(\text{higher order}\right)~.
\end{split}
\end{equation}
The higher-order terms are required by gauge invariance and supersymmetry. 

In general, the current multiplet~$\CJ$ encodes all couplings of the gauge theory to charged matter, as well as possible self-interactions due to higher-derivative terms, such as those indicated in~\eqref{action}. For simplicity, we will take all matter fields to reside in massless chiral multiplets.  Most of the results below only rely on general properties of the current multiplet~$\CJ$, e.g.~its supersymmetry transformations~\eqref{currsusyvar}, but do not depend on the detailed form of the interaction terms. Nevertheless, it is helpful to keep in mind the simplest theory in this class, which consists of a single massless, minimally coupled chiral multiplet of charge~$q$, with canonical kinetic terms and no superpotential or higher-derivative interactions.\footnote{~This theory is quantum mechanically anomalous. The anomaly can be cancelled by including additional chiral multiplets with suitable~$U(1)$ charge assignments.} In this theory, the operators in the current multiplet~$\CJ$ are given by
\begin{subequations}
\begin{align}
& \CK^B  = q \b \Phi \Phi~, \\
& \CK^F_\alpha  = q \sqrt 2 \, \b \Phi \Psi_\alpha~, \qquad \b \CK^F_\alphadot =  q \sqrt 2 \, \Phi \b \Psi_\alphadot~,\\
& \CJ_\mu  = q \left(i \b \Phi \, \overleftrightarrow {D_\mu} \Phi + \b \Psi \, \b \sigma_\mu \Psi\right)~.
\end{align}
\end{subequations}
Here~$\Phi, \Psi_\alpha$ are the propagating component fields in the chiral multiplet (their Hermitian conjugates~$\b \Phi, \b \Psi_\alphadot$ reside in an anti-chiral multiplet) and~$D_\mu = \d_\mu - i q A_\mu$ is the gauge-covariant derivative. In our conventions, the electric charge~$\mathscr E$ is given by
\begin{equation}
{\mathscr E} = \int d^3 x \, \CJ_0~,
\end{equation}
and the statement that~$\Phi, \Psi_\alpha$ both have charge~$q$ means that
\begin{equation}
\left[{\mathscr E}, \Phi(x)\right] = - q \Phi(x)~, \qquad \left[{\mathscr E}, \Psi_\alpha(x) \right] = - q \Psi_\alpha(x)~.
\end{equation}
This implies that a state~$|\Phi\rangle \sim \Phi(x) |0 \rangle$ created by~$\Phi(x)$ has charge~$-q$, since~${\mathscr E} |\Phi\rangle = - q |\Phi\rangle$.

\subsection{Scattering Amplitudes} 

We are interested in tree-level scattering amplitudes involving particles in the gauge and matter multiplets. An~$n$-point amplitude~$\CM_n$ is specified by~$n$ asymptotic one-particle states~$|f, p, s\rangle$, where~$p$ is the four-momentum of the particle and~$s$ is a spin or helicity label.  The label~$f$ denotes the particle type, i.e.~the field~$f(x)$ that creates the state, $|f\rangle \sim f(x) |0\rangle$.\footnote{~In this notation, $\langle f | \sim \langle 0 | \b f(x)$, while~$|\b f\rangle = \b f(x)|0\rangle$ and~$\langle \b f| \sim \langle 0 |f(x)$, where~$\b f(x)$ is the Hermitian conjugate of~$f(x)$.} For instance, we write~$|\CF, p, \pm\rangle$ for a photon of momentum~$p$ and helicity~$\pm1$. Similarly, $|\Lambda, p, -\rangle$ and~$|\b \Lambda, p, +\rangle$ are photini of momentum~$p$ and helicity~$-\half$ and~$+\half$, respectively. (See appendix~\ref{modeexp} for further details.) We normalize one-particle states as follows,
\begin{equation}
\langle f, p, s| f', p', s'\rangle = (2\pi)^3 \, (2 p^0) \, \delta_{f, f'}\delta_{s, s'} \delta^{(3)}(\vec p - \vec p\,')~.
\end{equation}
An~$n$-point amplitude~$\CM_n$ that describes~$m$ incoming particles~$\big\{ |f_i, p_i,s_{i}\rangle \big\}_{i = 1}^m$ scattering into~$n - m$ outgoing particles~$\big\{|f_i, p_i,s_{i}\rangle \big\}_{i = m+1}^n$ is given by 
\begin{equation}\label{ampdef}
\CM_n = \langle m+1 \, ; \,  \ldots \, ; \,  n  \, \big| \,\CS \, \big| \, 1 \, ; \,  \ldots \, ; \, m\rangle~, 
\end{equation}
where~$\CS$ is the scattering matrix and we have introduced the shorthand
\begin{equation}
|i\rangle = |f_i, p_i, s_i\rangle \qquad (i = 1, \ldots, n)~.
\end{equation}

\subsection{Soft Photon Theorem}
\label{softphoton}

Consider a scattering amplitude~$\CM_{n+1}^{\text{out}, \pm}$ involving an outgoing photon of momentum~$p_{n+1}$ and helicity~$\pm$, as well as~$n$ other asymptotic states (some of which may also be photons),
\begin{equation}
\CM_{n+1}^{\text{out}, \pm} = \left\langle m+1 \, ; \, \ldots \, ; \, n \, ; \, \CF, p_{n+1}, \pm \big| \, \CS \, \big| 1 \, ; \, \ldots \, ; \, m\right\rangle~.
\end{equation}
The leading behavior of this amplitude when the momentum of the photon is taken to zero, $p_{n+1} \rightarrow 0$, is governed by a universal soft theorem (see for instance~\cite{Weinberg:1995mt} and references therein),
\begin{equation}\label{soft-theorem}
\CM_{n+1}^{\text{out}, \pm}  \quad  {\longrightarrow} \quad e\left(\sum_{i=m+1}^n q_i \, {\ep^{(\pm)} \cdot p_i \over p_{n+1} \cdot p_i}  - \sum_{i=1}^m q_i \, {\ep^{(\pm)} \cdot p_i \over p_{n+1} \cdot p_i}\right) \CM_n~.
\end{equation}
Here~$\ep^{(\pm)}_\mu$ is the polarization vector of the soft photon, $q_i$ is the electric charge of particle~$i$, and~$\CM_n$ is the hard amplitude without the soft photon. The only assumptions that are needed to derive the soft photon theorem are Lorentz symmetry and gauge invariance. In order to set the stage for our discussion of supersymmetric soft theorems, we will now sketch a proof based on matrix elements of the electric current between asymptotic states. 

At tree level, the pole at~$p_{n+1} = 0$ on the right-hand side of~\eqref{soft-theorem} can only arise when an internal propagator goes on shell, which happens precisely when the soft photon attaches to one of the external lines. This is described by a single insertion of the interaction term~$- \CA^\mu \CJ_\mu \subset \SL_{\text{int}}$ in~\eqref{matintlag}. Factorizing on the propagators that go on shell when~$p_{n+1} \rightarrow 0$ then leads to 
\begin{equation} \label{photonsoftlimit}
\begin{split}
& \CM_{n+1}^{\text{out}, \pm}  \quad  \longrightarrow \quad - i  \bra{ \CF, p_{n+1} ,\pm } \A^\mu(0)  \ket{0} \times \cr
& \sum_{f,s}   \bigg(\sum_{i=m+1}^{n}  \frac{  - i  }{ 2  p_{n+1} \cdot p_{i} } \bra{i } \CJ_{\mu}(0) \ket{ f, p_{i} , s }  
\left\langle m+1 \, ; \, \ldots \, ; \, f, p_{i}, s \, ; \, \ldots \, ; n \big| \, \CS \, \big| 1\, ; \, \ldots \, ; \, m\right\rangle \cr
& ~~~~~~ -\sum_{i=1}^{m} \frac{  - i  }{ 2   p_{n+1}\cdot p_{i}} 
\left\langle m+1 \, ; \, \ldots \, ; n \big| \, \CS \, \big| 1\, ; \,\ldots \, ; \, f, p_{i}, s \, ; \, \ldots \, ; \, m\right\rangle  \bra{ f, p_{i} , s} \CJ_{\mu}(0) \ket{i  }  \bigg)~.
\end{split}
\end{equation}
It follows from the mode expansions in appendix~\ref{modeexp} that 
\begin{equation}
\begin{split}\label{Avev}
\bra{ \CF, p_{n+1} ,\pm } \A_\mu(0)  \ket{0} = e \ve^{(\pm)}_{\mu}~.
\end{split}
\end{equation}
Lorentz invariance and current conservation completely determine the matrix elements of the electric current $\CJ_\mu(x)$ between states of equal momentum (i.e.~in the forward limit) in terms of their electric charge (see for instance~\cite{Weinberg:1995mt}, chapter~10),
\begin{equation}\label{Jvev}
 \bra{f, p , s} \CJ_{\mu}(0) \ket{ f', p, s' } = - 2 q_f \, p_{\mu} \delta_{f f'} \delta_{s s'}~.
\end{equation}
Substituting~\eqref{Avev} and~\eqref{Jvev} into~\eqref{photonsoftlimit} establishes the soft photon theorem~\eqref{soft-theorem}. 

\subsection{Soft Photino Theorem}
\label{sec:softphotinothm}

In the supersymmetric case, we can study scattering amplitudes~$\CM^{\text{out},+}_{n+1}$ involving an outgoing photino~$\b \Lambda$ of momentum~$p_{n+1}$ and positive helicity, as well as~$n$ other hard particles,\footnote{~We add the photino to the out state by acting with its annihilation operator (see appendix~\ref{modeexp}) as follows, 
\begin{equation}\label{photinorule}
\big\langle m+1 \, ; \, \ldots \, ; \, n \, ; \, \b \Lambda, p_{n+1}, + \big| = \big\langle m+1 \, ; \, \ldots \, ; \, n \,  \big| a_{\b \Lambda, +}(p_{n+1})~.
\end{equation}
}
\begin{equation}
\CM^{\text{out},+}_{n+1} = \left\langle m+1 \, ; \, \ldots \, ; \, n \, ; \, \b \Lambda, p_{n+1}, + \big| \, \CS \, \big| 1 \, ; \, \ldots \, ; \, m\right\rangle~.
\end{equation}
In order for the amplitude to be non-zero, the total number of fermions involved in the scattering process (including the photino) must be even. We are interested in the leading behavior of this amplitude when the photino momentum is taken to zero, $p_{n+1} \rightarrow 0$. As in section~\ref{softphoton}, this arises from single insertions of the interaction terms~$-i \Lambda \CK^F + i \b \Lambda \, \b \CK^F \subset \SL_\text{int}$ in~\eqref{matintlag} that attach only to external lines. For a positive-helicity photino, insertions of~$-i \Lambda \CK^F$ do not contribute, since~$\bra{ \b \Lambda, p_s, + } \Lambda_\alpha(0)  \ket{0} = 0$. Therefore, the amplitude obeys the following soft theorem,
\begin{equation}
\begin{split}\label{photinosoftlimit}
& \CM^{\text{out},+}_{n+1} \quad  \longrightarrow \quad  -\bra{\b \Lambda, p_{n+1}, + } \b \Lambda_\alphadot (0)  \ket{0} \times \\
 & \sum_{f,s}   \bigg(\sum_{i=m+1}^{n}  \frac{  - i  }{ 2   p_{n+1} \cdot p_{i}}  (-1)^{\sigma_i} \bra{i } \b \CK^{F\alphadot}(0) \ket{ f, p_{i} , s }  
\left\langle m+1 \, ; \, \ldots \, ; \, f, p_{i}, s \, ; \, \ldots \, ; n \big| \, \CS \, \big| 1\, ; \, \ldots \, ; \, m\right\rangle \cr
& -\sum_{i=1}^{m} \frac{  - i  }{ 2 p_{n+1} \cdot p_{i} } 
(-1)^{\sigma_i} \left\langle m+1 \, ; \, \ldots \, ; n \big| \, \CS \, \big| 1\, ; \,\ldots \, ; \, f, p_{i}, s \, ; \, \ldots \, ; \, m\right\rangle  \bra{ f, p_{i} , s} \b \CK^{F\alphadot}(0) \ket{i  }  \bigg)~.
\end{split}
\end{equation}
Here~$(-1)^{\sigma_i}$ is a fermion sign factor that comes from anticommuting~$\b \CK^{F\alphadot}$ across multi-particle states.\footnote{~For an out state~$\langle m+1 \, ; \, \ldots \, ; \, i \,  ; \,  \ldots \, ; \, n  |$ we define~$\sigma_i$ to be the number of fermionic states in positions~$i+1$ through~$n$. For an in state~$|1 \,; \,  \ldots \, ; \, i \, ;  \,  \ldots \, ;  \, m  \rangle$ we define it to be the number of fermionic states in positions~$1$ through~$i-1$.}

The photino wavefunction is given by (see appendix~\ref{modeexp})
\begin{equation}
\bra{\b \Lambda, p_{n+1}, + } \b \Lambda_\alphadot (0)  \ket{0} = e \b \eta_\alphadot(p_{n+1})~.
\end{equation}
Here~$\b \eta_\alphadot(p)$ is a standard spinor-helicity variable corresponding to a null momentum~$p$,\footnote{~In terms of the usual angle and square brackets, we have~$\eta_\alpha(p_i) = i\rangle_\alpha$ and~$\b \eta^\alphadot(p_i) =  i ]^\alphadot$. }
\begin{equation}\label{spinorhelvar}
p_\mu \sigma^\mu_{\alpha\alphadot} = \eta_\alpha(p) \b \eta_\alphadot(p)~.
\end{equation}

We must now evaluate the matrix elements of the fermionic operator~$\b \CK^{F \alphadot}$ between single-particle states, in the forward limit. In general, the matrix elements of such an operator may be model-dependent. However, $\b \CK^{F \alphadot}$ resides in the same supermultiplet~\eqref{jmult} as the conserved electric current~$\CJ_\mu$, whose forward matrix elements are universal, as discussed around~\eqref{Jvev}. Explicitly, we can evaluate the following commutation relation from~\eqref{currsusyvar},
\begin{equation}
\big\{Q_\alpha, \b \CK^F_\alphadot\big\} = i \sigma^\mu_{\alpha\alphadot} \left(\CJ_\mu - i \d_\mu \CK^B\right)~,  
\end{equation}
between single-particle states in the forward limit, where we can drop the total derivative~$\d_\mu \CK^B$. Using~\eqref{Jvev} then leads to
\begin{equation}\label{qkbme}
 \bra{f, p , s} \big\{Q_\alpha, \b \CK^F_\alphadot\big\}\ket{ f', p, s' } = - 2i q_f \, p_{\mu} \sigma^\mu_{\alpha\alphadot}  \, \delta_{f f'} \delta_{s s'}~.
\end{equation}
The appearance of~$\delta_{f f'}$ on the right-hand side shows that only single-particle states that reside in the same supermultiplet can can lead to non-vanishing matrix elements for~$\b \CK^{F}_\alphadot$. When the supercharges act on the left or the right, they lead to other states in this supermultiplet, in a way that is completely determined by representation theory. This can be used to derive all matrix elements of~$\b \CK^F_\alphadot$ between massless or massive single-particle states of arbitrary spin.

Here we explicitly work this out for a massless chiral multiplet~$\Phi, \Psi_\alpha$ of charge~$q$, and its conjugate anti-chiral multiplet~$\b \Phi, \b \Psi_\alphadot$ of charge~$-q$. The relevant single-particle states are
\begin{equation}
|\Phi, p\rangle~, \quad |\Psi, p, -\rangle \qquad \text{and} \qquad |\b \Phi, p\rangle~, \quad | \b \Psi, p, +\rangle~. 
\end{equation}
On these states, the supersymmetry algebra is represented as follows,\footnote{~This follows from the non-vanishing commutation relations for a free chiral multiplet, $[Q_\alpha, \Phi] = i \sqrt 2 \Psi_\alpha$ and~$\{\b Q_\alphadot, \Psi_\alpha\} = \sqrt 2 \sigma^\mu_{\alpha\alphadot} \d_\mu \Phi$, as well as the mode expansions in appendix~\ref{modeexp}.}
\begin{align}\label{susychimultstate}
& \b Q_\alphadot |\Phi, p \rangle = 0~, & & Q_\alpha |\Phi, p\rangle = \sqrt 2 i  \, \eta_\alpha(p) |\Psi, p, -\rangle~,\cr
& \b Q_\alphadot |\Psi, p, -\rangle = - \sqrt 2 i \, \b \eta_\alphadot(p) |\Phi, p\rangle~, & & Q_\alpha |\Psi, p, -\rangle = 0~.
\end{align}
The action of the supercharges on the conjugate anti-chiral states is obtained by exchanging~$Q_\alpha \leftrightarrow \b Q_\alphadot$, $|\Phi, p\rangle \leftrightarrow |\b \Phi, p\rangle$, $\eta_\a(p) \leftrightarrow \b\eta_{\alphadot}(p)$, and~$|\Psi, p, -\rangle \leftrightarrow |\b \Psi, p, +\rangle$.

We can now implement the procedure described after~\eqref{qkbme} to obtain all non-vanishing matrix elements of~$\b \CK^F_\alphadot$,
\begin{equation}\label{kbmatel}
\langle \Phi, p| \b \CK_\alphadot^F(0) | \Psi, p, -\rangle =  \langle \b \Psi, p, +| \b \CK_\alphadot^F(0) | \b \Phi, p\rangle = \sqrt 2  q \b \eta_\alphadot(p)~. 
\end{equation}
Substituting into~\eqref{photinosoftlimit}, we obtain the final form of the soft photino theorem,
\begin{equation}\label{softphotino}
\CM^{\text{out},+}_{n+1} ~~~ \longrightarrow ~~~ { \sqrt 2 i e} \sum_{i = m+1}^n \,  {q_i \over \eta(p_{n+1}) \eta(p_i)} \left({\mathscr F}_i \CM_n\right) - { \sqrt 2 i e} \sum_{i = 1}^m \, {q_i \over \eta(p_{n+1}) \eta(p_i)} \left({\mathscr F}_i \CM_n\right)~. 
\end{equation}
Here the~$q_i$ are the electric charges of the asymptotic states. The~$n$-particle amplitude~$\CM_n$ is obtained from~$\CM_{n+1}^{\text{out},+}$ by deleting the photino, but since it has an odd number of fermion external states, it vanishes. The non-vanishing~$n$-point amplitude~$ {\mathscr F}_i \CM_n$ is obtained from~$\CM_n$ by acting on the~$i$th single-particle state with a fermionic operator~$\mathscr F$, which satisfies
\begin{equation}
\begin{split}\label{scriptFdef}
&  \langle \Phi, p|{\mathscr F} = - \langle \Psi, p, -|~, \qquad \langle \b \Psi, p, +| {\mathscr F}  = \langle \b \Phi, p|~, \\
&  {\mathscr F}|\b \Phi, p\rangle = |\b \Psi, p, +\rangle~, \qquad  {\mathscr F}|\Psi, p, -\rangle = - |\Phi, p\rangle~.
\end{split}
\end{equation} 
The action of~$\mathscr F$ on all other single-particle states vanishes, and we take~$\mathscr F$ to act from the right on out states and from the left on in states.  Since~$\mathscr F$ is a fermionic operator, it picks up a sign whenever it moves past another fermionic operator or state. This accounts for the factors~$(-1)^{\sigma_i}$ in~\eqref{photinosoftlimit}. 

So far we have only discussed an outgoing soft photino of positive helicity. The negative helicity case can similarly be shown to satisfy
\begin{equation}\label{softphotinoneg}
\CM^{\text{out},-}_{n+1} \longrightarrow  - { \sqrt 2 i e} \sum_{i = m+1}^n \,  {q_i \over \b \eta(p_{n+1}) \b\eta(p_i)} \left({\mathscr F}^\dagger_i \CM_n\right) + { \sqrt 2 i e} \sum_{i = 1}^m \, {q_i \over \b\eta(p_{n+1}) \b \eta(p_i)} \left({\mathscr F}^\dagger_i \CM_n\right)~,
\end{equation}
where the fermionic operator~${\mathscr F}^\dagger$ is the Hermitian conjugate of the operator~$\mathscr F$ defined in~\eqref{scriptFdef}. Finally, the soft theorems for ingoing photini can be obtained from~\eqref{softphotino} and~\eqref{softphotinoneg} by crossing symmetry.

\section{Asymptotic Symmetries}

\label{sec:asymni}

\subsection{Kinematics}
\label{sec:prelim}

We are studying scattering from past to future null infinity in flat Minkowski space, $ds^2 = \eta_{ab} d y^a dy^b$ with~$\eta_{ab} = -$+++. In order to discuss physics at null infinity, it is convenient to choose a different set of coordinates~$x^\mu = (u,r,z,\bz)$, in which the Minkowski metric takes the form
\begin{equation}
\begin{split}
ds^{2} = - du dr + r^{2} dz d\bz~.
\end{split}
\end{equation}
These coordinates degenerate at~$r = 0$, but this will not affect our discussion of the  asymptotic regions at large~$r$. The transformation to conventional flat coordinates~$y^a$ is given by\footnote{~The~$(u, r, z, \b z)$ coordinates are simply related to other useful coordinate systems. For instance, they can be obtained from standard retarded coordinates~$(u', r', z', \b z')$, in which~$ds^2 = - du'^2 - 2 du' dr' + {4 r'^2 \over (1 + |z'|^2)^2} dz' d \b z'$, by setting~$u' = c u$, $r' = {1 \over 2 c} r$, $z' = c z$ and taking the constant~$c \rightarrow 0^+$. By further redefining~$r \rightarrow - {1\over r}$, we obtain the coordinates that were used in~\cite{Hofman:2008ar} to study energy and charge correlators at null infinity.}
\begin{equation}\label{yxcoords}
\begin{split}
y^0 &= \frac{1}{2} \left(u + r  \left( 1+|z|^2\right) \right)~, \\
y^1 &= \frac{r}{2} \left( z + \bz \right)~, \\
y^2 &= - \frac{ir}{2} \left( z - \bz \right)~, \\
y^3 &= - \frac{1}{2} \left( u - r  \left( 1 - |z|^2 \right)  \right)~. 
\end{split}
\end{equation}
The null coordinates~$-\infty < u, r < \infty$ are real, while~$z \in \C$ is complex. Future and past null infinity~$\ci^+$ and~$\ci^-$ are located at~$r \rightarrow +\infty$ and~$r \rightarrow - \infty$, respectively (see appendix~\ref{app:conventions} for additional details). Both have topology~$\R \times S^2$, where~$\R$ is a null direction parametrized by~$u$, while the~$S^2$ is spacelike. The complex variable~$z$ is an angular coordinate, which covers all of~$S^2$, except one point that will not be important for us. Note that a light ray traversing Minkowski space hits the same angular coordinate~$z$ at both~$\ci^+$ and~$\ci^-$, i.e.~points with the same angular coordinate are identified by the antipodal map. 

The future and past boundaries of~$\ci^+$ are at~$u \rightarrow \pm \infty$ and will be denoted by~$\ci^+_\pm$. Given a field~$\CO(u, z, \b z)$ on~$\ci^+$, its boundary values at~$\ci^+_\pm$ are given by the following limits, assuming they exist,
\begin{equation}
\CO|_{\ci^+_\pm} = \lim_{u \rightarrow \pm \infty} \CO(u, z, \b z)~.
\end{equation}
Similarly, the boundaries of~$\ci^-$ at~$u \rightarrow \pm \infty$ are~$\ci^-_\pm$, and the corresponding boundary values of a field~$\CO^-(u, z, \b z)$ on~$\ci^-$ are denoted by~$\CO^-|_{\ci^-_\pm}$. We will generally label fields on~$\ci^-$ using the same symbols as the corresponding fields on~$\ci^+$, but with an extra superscript~$-$\,. Spatial infinity is pinched between~$\ci^+_-$ and~$\ci^-_+$. 

In order to discuss spinors, it is convenient to choose the vielbein
\begin{equation}\label{vielbein}
e^a_\mu dx^\mu = dy^a = {\d y^a \over \d x^\mu} \, dx^\mu~,
\end{equation}
with~$y^a(x^\mu)$ as in~\eqref{yxcoords}, since it leads to a vanishing spin connection. Therefore, covariant derivatives of spinors coincide with ordinary partial derivatives. This makes it straightforward to adapt standard flat-space formulas (e.g.~supersymmetry transformations) to~$(u, r, z, \b z)$ coordinates.  Another simplification comes from using a helicity basis for spinors. In the frame~\eqref{vielbein}, we define the following linearly independent, commuting, left-handed basis spinors, 
\begin{equation}\label{xidef}
\begin{split}
\xi^{(+)}_\a(z) = \left(\begin{array}{c} 1 \\ z \end{array} \right) ~, \qquad \xi^{(-)}_\a = \left(\begin{array}{c} 0 \\ 1 \end{array} \right)~,
\end{split}
\end{equation}
as well as their right-handed complex conjugates~${\bar \xi}^{(+)}_\da(\b z)$ and ${\bar \xi}^{(-)}_\da$. (We will often suppress the dependence of~$\xi_\alpha^{(+)}$ and~$\b \xi_\alphadot^{(+)}$ on~$z$ and~$\b z$.) They satisfy 
\begin{equation}
\xi^{(+)\alpha} \xi^{(-)}_\alpha = -1~, \qquad \b \xi^{(+)}_\alphadot \b \xi^{(-)\alphadot} = 1~.
\end{equation}
Note that~$\xi^{(-)}_\alpha$ is (covariantly) constant, while~$\d_z \xi_\alpha^{(+)} = \xi_\alpha^{(-)}$ and~$\d_{\b z} \xi_\alpha^{(+)} = 0$. All~$\sigma$-matrices can be expressed as bilinears in~$\xi^{(\pm)}_\alpha$ and~$\b \xi^{(\pm)}_\alphadot$ (see appendix~\ref{app:conventions}). 

The spinors~$\xi^{(+)}_\alpha$, $\b \xi^{(+)}_\alphadot$ are position space analogues of the momentum space spinor-helicity variables~$\eta_\alpha(p), \b \eta_\alphadot(p)$ introduced in~\eqref{spinorhelvar}. In order to see this, we parametrize the null momentum~$p^\mu$ as follows,
\begin{equation}\label{pdef}
\begin{split}
p^\mu \d_\mu = 2 \omega \d_r  = \omega \left( \left(1 + |z|^2\right)\d_{y^0} + \left(z + \bz\right)\d_{y^1}  - i \left(z - \bz\right)\d_{y^2} + \left(1 - |z|^2\right)\d_{y^3} \right)~.
\end{split}
\end{equation}
Here~$\omega > 0$ has units of energy, but it only coincides with the energy~$p^0$ when~$z = 0$. It follows from~\eqref{pdef} that
\begin{equation}
p^\mu \sigma_{\mu \alpha\alphadot} =  2 \omega \xi^{(+)}_\alpha(z) \b \xi^{(+)}_\alphadot(\b z)~,
\end{equation}
so that the spinor-helicity variables defined in~\eqref{spinorhelvar} are given by\footnote{~The relation in~\eqref{spinorhelvar} only defines~$\eta_\alpha(p)$ up to a little group phase, which is fixed by~\eqref{etaxirel}.}
\begin{equation}\label{etaxirel}
\eta_\alpha(p) = \sqrt{2 \omega} \xi^{(+)}_\alpha(z)~, \qquad \b \eta_\alphadot(p) = \sqrt{2 \omega} \, \b \xi^{(+)}_\alphadot(\b z)~. 
\end{equation}

The Lorentz group~$SL(2, \C)$ is isomorphic to the global conformal group in two Euclidean dimensions, which acts on the asymptotic~$S^2$ parametrized by~$z$. Infinitesimal Lorentz transformations are therefore parametrized by two-dimensional holomorphic vector fields~$Y^z = a + b z + z^2$ with~$a, b, c \in \C$. The corresponding Lorentz transformation~$L_Y$ is given by\begin{equation}\label{lordef}
\begin{split}
L^\mu_Y \p_\mu = \frac{1}{2} \left( \p_z Y^z + \p_\bz Y^\bz \right)  \left( u \p_u - r \p_r \right) + \left( Y^z - \frac{u}{2r}  \p_\bz^2 Y^\bz   \right) \p_z + \left( Y^\bz - \frac{u}{2r}  \p_z^2 Y^z   \right) \p_\bz~.
\end{split}
\end{equation}
The~$SO(2)$ rotation that stabilizes the null vector field~$p^\mu$ in~\eqref{pdef} is generated by the following infinitesimal Lorentz transformation,
\begin{equation}
J = i \left(z \d_z - \b z \d_{\b z}\right)~.
\end{equation}
Under such a rotation, the spinors in~\eqref{xidef} transform as follows,\footnote{~See appendix~\ref{app:conventions} for the action of the Lie derivative on spinors.}
 \begin{equation}\label{xihelicity}
\CL_{J} \xi^{(\pm)}_\alpha = \pm {i \over 2} \xi_\alpha^{(\pm)}~, \qquad \CL_{J} \b \xi^{(\pm)}_\alphadot = \mp {i \over 2} \b \xi_\alphadot^{(\pm)}~.
\end{equation}
This shows that the spinors~$\xi_\alpha^{(\pm)}$ and~$\b \xi_\alphadot^{(\mp)}$ have definite helicity~$\pm \half$, and hence they are chiral spinors on the asymptotic~$S^2$ parametrized by~$z$. 

Given a field~$F_{(\alpha_1 \cdots \alpha_{2 j}) (\betadot_{1} \cdots \betadot_{2 \b j})}$ that transforms in the~$(j, \b j)$ representation of the Lorentz group~$SL(2, \C) = SU(2) \times \b{SU(2)}$, with~$j, \b j \in \half \Z$, it is natural to expand it in terms of the helicity eigenspinors~$\xi_\alpha^{(\pm)}, \b \xi_\betadot^{(\pm)}$, 
\begin{equation}\label{helexp}
F_{(\alpha_1 \cdots \alpha_{2 j}) (\betadot_{1} \cdots \betadot_{2 \b j})} = \sum_{m = - j}^m \sum_{\b m = - \b j}^{\b j} \, \xi^{(+)}_{(\alpha_1} \cdots \xi^{(+)}_{\alpha_{j+m}} \xi^{(-)}_{\alpha_{j+m + 1}} \cdots \xi^{(-)}_{\alpha_{2j)}} \b \xi^{(+)}_{(\betadot_1} \cdots \b \xi^{(+)}_{\betadot_{\b j+\b m}} \b \xi^{(-)}_{\betadot_{\b j+\b m + 1}} \cdots \b \xi^{(-)}_{\betadot_{2\b j)}} F_{(m, \b m)}~.
\end{equation}
As we will discuss in examples below, the coefficient fields~$F_{(m, \b m)}$ obey simple falloff conditions near null infinity. In order to state these conditions, we need to introduce a conformal scaling dimension~$\Delta$ for the field~$F_{(\alpha_1 \cdots \alpha_{2 j}) (\betadot_{1} \cdots \betadot_{2 \b j})}$, even though the theory under consideration need not be conformally invariant. Nevertheless, we expect its long-distance behavior near null infinity to be governed by a conformally invariant IR fixed point, and we take~$\Delta$ to be the scaling dimension of~$F_{(\alpha_1 \cdots \alpha_{2 j}) (\betadot_{1} \cdots \betadot_{2 \b j})}$ at that fixed point. In this paper, we are interested in abelian gauge theories, which are IR free. In these theories~$\Delta$ coincides with the engineering dimensions of the field~$F_{(\alpha_1 \cdots \alpha_{2 j}) (\betadot_{1} \cdots \betadot_{2 \b j})}$. The behavior of the coefficient field~$F_{(m, \b m)}$ near~$\ci^\pm$ (i.e.~for~$r \rightarrow \pm \infty$) is governed by~$\Delta$ and its Lorentz quantum numbers~$m, \b m$, \begin{equation}\label{rfalloff}
F_{(m, \b m)}(r, u, z, \b z) = \CO\left({1 \over |r|^{\Delta - m - \b m}}\right)~.
\end{equation}
The quantity~$\Delta - m - \b m$ is known as the collinear twist: it is the eigenvalue of the conformal generator~$D + M$, which stabilizes the null vector field~$p^\mu$.\footnote{~Here~$D = u \d_u + r \d_r$ is a dilatation, which satisfies~$[D, p^\mu \d_\mu] = - p^\mu \d_\mu$, and~$M = u \d_u  - r \d_r + z \d_z + \b z \d_{\b z}$ is a boost along~$p^\mu$, which satisfies~$[M, p^\mu \d_\mu] = p^\mu \d_\mu$.} As a simple example, consider an IR free, massless scalar field~$\Phi$ of scaling dimension~$\Delta = 1$. According to~\eqref{rfalloff}, its asymptotic expansion near~$\ci^+$ is
\begin{equation}\label{scalarlarger}
\Phi(u, r, z, \b z) = {1 \over r} \phi(u, z, \b z) + \CO(r^{-2})~,
\end{equation}
and similarly near~$\ci^-$, with~$\phi \rightarrow \phi^-$.

\subsection{Photon Asymptotics and Bosonic Asymptotic Symmetries}\label{sec:photonasymptotics}

The photon is described by an anti-symmetric field strength~$\CF_{\mu\nu}$, whose IR scaling dimension is~$\Delta_\CF = 2$. It decomposes into self-dual and anti-self-dual parts~$\CF_{\mu\nu}^\text{SD}$ and~$\CF_{\mu\nu}^\text{ASD}$, which transform as~$(1,0)$ and~$(0,1)$ representations of the Lorentz group. According to~\eqref{rfalloff}, the different components of~$\CF^\text{SD}_{\mu\nu}$ behave as follows near~$\ci^\pm$,
\begin{equation}\label{flarger}
\begin{split}
\CF^\text{SD}_{(1,0)} & \sim {1 \over r} \CF_{uz} = \CO\left({1 \over r}\right)~, \\ \CF^\text{SD}_{(0,0)} & \sim \CF_{ur} - {1 \over r^2} \CF_{z \b z} = \CO\left({1 \over r^2}\right)~, \\
\CF_{(-1,0)}^{SD} & \sim {1 \over r} \CF_{r \b z} = \CO\left({1 \over r^3}\right)~.
\end{split}
\end{equation}
This is consistent with the following asymptotic expansion for the gauge field~$\CA_\mu$ near~$\ci^+$, 
\begin{equation}
\begin{split}\label{photonlarger}
 \A_{u} (u,r,z,\bz) &= \frac{1}{r} \, A_{u} (u,z,\bz) + \Or(r^{-2})~, \\
 \A_{r}  (u,r,z,\bz) &= \frac{1}{r^{2}} \, A_{r} (u,z,\bz) + \Or(r^{{-3}})~, \\
 \A_{z} (u,r,z,\bz)  &=  A_{z} (u,z,\bz) + \Or(r^{-1})~.
\end{split}
\end{equation}
These falloffs were postulated in recent discussions of the classical scattering problem for abelian gauge fields~\cite{as,hmps,He:2015zea,Campiglia:2015qka,Kapec:2015ena,Strominger:2015bla}, where they were shown to satisfy various physical requirements, e.g.~they lead to a finite energy energy flux through~$\ci^+$. 

The dynamics of~$\CA_\mu$ is governed by Maxwell's equations, as determined by the Lagrangian~\eqref{action} and~\eqref{matintlag}, 
\begin{equation}
\begin{split}\label{eom1}
\nabla^{\mu} \F_{\mu\nu} = e^{2} \CJ_{\nu}~.
\end{split}
\end{equation}
Substituting~\eqref{flarger} or~\eqref{photonlarger} into this equation leads to the following asymptotic expansion for the electric current~$\CJ_\mu$ near~$\ci^+$,
\begin{equation}
\begin{split}\label{currlarger}
\CJ_{u} (u,r,z,\bz) &= \frac{1}{r^2} \, j_{u} (u,z,\bz) + \Or(r^{-3})~, \\
 \CJ_{r}  (u,r,z,\bz) &= \frac{1}{r^{4}} \, j_{r} (u,z,\bz) + \Or(r^{{-5}})~, \\
 \CJ_{z} (u,r,z,\bz)  &=  {1 \over r^2} \, j_{z} (u,z,\bz) + \Or(r^{-3})~.
\end{split}
\end{equation}
This is consistent with the general prescription~\eqref{rfalloff} and the fact that~$\CJ_\mu$ should be an operator of IR scaling dimension~$\Delta_\CJ = 3$ in the~$(\half, \half)$ representation of the Lorentz group.\footnote{~The coefficients of~$\CJ_\mu$ in the helicity expansion~\eqref{helexp} are
\begin{equation}
\CJ_{(\half, \half)} \sim \CJ_u~, \qquad \CJ_{(- \half, \half)} \sim {1 \over r} \CJ_{\b z}~, \qquad \CJ_{(\half, -\half)} \sim {1 \over r} \CJ_z~, \qquad \CJ_{(-\half, -\half)} \sim \CJ_r~.
\end{equation}
}

Substituting the asymptotic expansions~\eqref{photonlarger} and~\eqref{currlarger} into the equations of motion~\eqref{eom1} and expanding in powers of~$1 \over r$ leads to a system of equations that determines (together with suitable boundary conditions at~$\ci^+_\pm$) the bulk gauge field~$\CA_\mu$ in terms of the boundary data~$A_z(u, z, \b z)$ and the current~$\CJ_\mu$, up to gauge transformations that vanish at~$\ci^+$ and hence do not act on physical states. In particular, we obtain the leading constraint equation on~$\ci^{+}$,
\begin{equation}\label{constrainteq}
\begin{split}
\p_{u}  F_{ur}  = \p_{u}   \left( \p_{z} A_{\bz } + \p_{\bz} A_{z} \right)  + \frac{e^{2}}{2} j_{u} ~,
\end{split}
\end{equation}
where~$F_{ur}$ is the leading term in the asymptotic expansion of~$\CF_{ur}$,
\begin{equation}
\CF_{ur} = {1 \over r^2} F_{ur} + \CO(r^{-3})~, \qquad F_{ur} = A_{u} +  \p_{u} A_{r}~.
\end{equation}

The boundary data on~$\ci^{+}$ is acted on by an infinite number of asymptotic symmetries, parametrized by a function~$\ep(z, \b z)$. The corresponding charges, which generate the symmetries, are denoted by~${\mathscr E}[\ep]$,
\begin{equation}\label{bcomms}
\begin{split}
\big[{\mathscr E}[\ep], A_z(u, z, \b z)\big] = i \d_z \ep(z, \b z)~, \qquad \big[{\mathscr E}[\ep], f_q(u, z, \b z)\big] = - q \ep(z, \b z) f_q(u, z, \b z)~.
\end{split}
\end{equation}
Here~$f_q(u, z, \b z)$ denotes an arbitrary boundary field of charge~$q$.  For real $\ep$, these symmetries can be viewed as large gauge transformations that do not vanish on~$\ci^+$. It follows from~\eqref{bcomms} that~${\mathscr E}[\ep = 1]$ is the electric charge, which is linearly realized on the fields. For non-constant~$\ep(z, \b z)$, the~${\mathscr E}[\ep]$ are spontaneously broken, since the~$u$-independent modes of the photon shift inhomogenously. These modes describe soft, zero-momentum photons (see appendix~\ref{modeexp}) and can be interpreted as Goldstone bosons for the spontaneously broken charges~\cite{hmps}. 

In theories with magnetic charges, there is a second set of conserved charges and asymptotic symmetries that can similarly be viewed as large magnetic gauge transformations acting on a dual magnetic vector potential \cite{Strominger:2015bla}. These symmetries can be incorporated by simply allowing~$\ep(z, \b z)$ in~\eqref{bcomms} to be a complex-valued function. For simplicity, we will restrict the discussion below to theories that only contain electric charges, in which the magnetic symmetries act trivially.  

In the absence of magnetic charges, it was shown in~\cite{hmps} that the~${\mathscr E}[\ep]$ can be written as integrals over~$\ci^+_-$,
\begin{equation}\label{bcharge}
{\mathscr E}[\ep] = - \frac{1}{e^{2}} \int d^{2} z \, \ve(z, \b z) \,   F_{ur}  \big|_{\ci^{+}_{-}}~.
\end{equation} 
By rewriting this as an integral of~$\d_u F_{ur}$ over all of~$\ci^+$ and using the constraint equation~\eqref{constrainteq}, the charges can be expressed as a sum of hard and soft contributions,
\begin{equation}\label{ehs}
\begin{split}
& {\mathscr E}[\ep] = {\mathscr E}^\text{h}[\ep] + {\mathscr E}^\text{s}[\ep]~,
\end{split}
\end{equation}
where the hard charge is given by
\begin{equation}\label{ehard}
{\mathscr E}^\text{h}[\ep] = \half \int du d^2 z \, \ep(z, \b z) j_u - {1 \over e^2} \int d^2 z \, \ep(z, \b z) F_{ur}\big|_{\ci^+_+}~.
\end{equation}
The first term arises from massless charged particles passing through~$\ci^+$~\cite{hmps}, while the second one is due to massive charged particles passing through future timelike infinity~\cite{Campiglia:2015qka,Kapec:2015ena}. In this paper we only consider massless particles, and hence we set~$F_{ur}|_{\ci^+_+} = 0$.  We also assume that the electric current~$j_u$ vanishes at~$\ci^+_\pm$, so that the flow of charge through~$\ci^+$ is, in a suitable sense, bounded.  (Below we will impose the same restriction on the other sources in the current multiplet~$\CJ$.) For instance, this is the case for a scattering process involving a finite number of charged particles. 

The soft charge in~\eqref{ehs} is given by\begin{equation}\label{esoft}
 {\mathscr E}^\text{s}[\ep]  = -{1 \over 4 \pi} \int d^2 z \, \ep(z, \b z) \left(\d_{\b z} j^\text{s}_z + \d_z j^\text{s}_{\b z}\right)~, \qquad j^\text{s}_z(z, \b z) = -{4 \pi \over e^2} \int du \, \d_u A_z~.
\end{equation}
Here we have defined the soft photon current~$j^\text{s}_z$. (The normalization is as in~\cite{He:2015zea}.) Under a Lorentz transformation~\eqref{lordef}, it transforms as
\begin{equation}\label{jsofttrans}
\delta_Y j^\text{s}_z = \left(\d_z Y^z + Y^z \d_z + Y^{\b z} \d_z\right) j^\text{s}_z~,
\end{equation}
up to contributions from~$\ci^+_\pm$ that vanish whenever the field~$A_z$ has well-defined boundary values~$A_z |_{\ci^+_\pm}$. The transformation~\eqref{jsofttrans} implies that~$j^\text{s}_z$ transforms as a two-dimensional field with~$SL(2, \C)$ conformal weights~$h = 1$ and~$\b h = 0$, which are appropriate for a left-moving Kac-Moody current.

It is straightforward to repeat the preceding discussion near~$\ci^-$. In particular, the charges~${\mathscr E}^-[\ep]$ that generate the asymptotic symmetries at~$\ci^-$ are given by 
\begin{equation}\label{bminuscharge}
\begin{split}
{\mathscr E}^-[\ep] = - \frac{1}{e^{2} } \int d^{2} z \, \ep(z, \b z)   F^-_{ur}   \big|_{\ci^{-}_{+}}~.
\end{split}
\end{equation}
Here~$\displaystyle F^-_{ur} = \lim_{r \rightarrow -\infty} r^2 \CF_{ur}$ is the boundary value of the electric field on~$\ci^-$. The fields on~$\ci^+$ and~$\ci^-$ satisfy Lorentz-invariant matching conditions at spatial infinity. Since we are considering gauge theories without magnetic charges, we follow~\cite{hmps} and demand that 
\begin{subequations}\label{Amatchcond}
\begin{align}
& A_{z} \big|_{\ci^{+}_{-}} = A^-_{z} \big|_{\ci^{-}_{+}}~,\label{Amatchcondi}\\
& F_{ur} \big|_{\ci^{+}_{-}} = F^-_{ur} \big|_{\ci^{-}_{+}}~.  \label{Amatchcondii}
\end{align}
\end{subequations}
Comparing with~\eqref{bcharge} and~\eqref{bminuscharge} leads to the conservation law
\begin{equation}
\begin{split} \label{chargeconservation}
{\mathscr E}[\ep]  = {\mathscr E}^-[\ep]~. 
\end{split}
\end{equation}
For~$\ep = 1$ this reduces to electric charge conservation. Semiclassically, the conservation law~\eqref{chargeconservation} amounts to a Ward identity for the tree-level~$\CS$-matrix,
\begin{equation}
\begin{split}\label{wardidphoton}
{\mathscr E}[\ep]  {\cal S} - {\cal S}{\mathscr E}^-[\ep] = 0~.
\end{split}
\end{equation}
It was shown in~\cite{hmps} that this Ward identity, evaluated between in and out scattering states, reproduces the soft photon theorem~\eqref{soft-theorem}. Moreover, it was shown in~\cite{as, He:2015zea} that both can be interpreted as a Kac-Moody Ward identity for the two-dimensional soft current~$j_z^\text{s} - {j_z^\text{s}}^-$, where~$j_z^\text{s}$ was defined in~\eqref{esoft} and~${j_z^\text{s}}^-$ is its counterpart on~$\ci^-$. 

\subsection{Photino Asymptotics}\label{sec:photinoasymptotics}

The dynamics of the photino field~$\Lambda_\alpha$ is governed by the following equations of motion, which can be derived from~\eqref{action} and \eqref{matintlag},
\begin{equation}\label{gauginoeom}
\begin{split}
{\bar \sigma}^{\mu \alphadot\alpha} \d_\mu \Lambda_{\alpha} = e^2 \bar \CK^{F\alphadot}~.
\end{split}
\end{equation}
As in~\eqref{helexp}, we project the photino~$\Lambda_\alpha$ and the fermionic source~$ \CK^F_\alpha$, both of which transform as~$(\half, 0)$ under the Lorentz group, onto the basis spinors~$\xi^{(\pm)}_\alpha$ defined in~\eqref{xidef},
\begin{align}
\begin{split}\label{lambdaexpansion}
\Lambda_{\a} &= \L_{(\half, 0)} \, \xi_{\a}^{(+)} + \Lambda_{(- \half, 0)} \, \xi_{\a}^{(-)}~, \qquad	 \CK_{\a}^F = \CK^F_{(\half, 0)} \, \xi_{\a}^{(+)} + \CK^F_{(- \half, 0)} \, \xi_{\a}^{(-)}~.
\end{split}
\end{align}
Following the discussion around~\eqref{rfalloff}, the leading large-$r$ behavior of the coefficient fields $\Lambda_{(\pm \half, 0)}$ and~$\CK^F_{(\pm \half, 0)}$ is determined by their conformal scaling dimensions: $\Lambda_\alpha$ is a free fermion of scaling dimension~$\Delta_\Lambda = {3 \over 2}$, while~$\CK^F_\alpha$ has dimension~$\Delta_{\CK^F} = {5 \over 2}$. Applying~\eqref{rfalloff} then leads to the following asymptotic expansions near~$\ci^+$, 
\begin{equation}
\begin{split}\label{photinolarger}
\Lambda_{(\half, 0)} (u,r,z,\bz) &=  \frac{1}{r} \lambda_{(+)}  (u,z,\bz) + \Or(r^{-2}) ~ , \\
\Lambda_{(-\half, 0)}(u,r,z,\bz) &=  \frac{1}{r^{2}} \lambda_{(-)}  (u,z,\bz) + \Or(r^{-3}) ~,
\end{split}
\end{equation}
and
\begin{equation}\label{kappalarger}
\begin{split}
\CK^F_{(\half, 0)}(u,r,z,\bz) &=  \frac{1}{r^{2}} k_{(+)} (u,z,\bz) + \Or(r^{-3})~, \\
\CK^F_{(-\half, 0)} (u,r,z,\bz) &=  \frac{1}{r^{3}} k_{(-)}  (u,z,\bz) + \Or(r^{-4})~.
\end{split}
\end{equation}
As for the photon, the equations of motion~\eqref{gauginoeom} determine the bulk field~$\Lambda_\alpha$ from the boundary data~$\lambda_{(+)}$ and the source~$\CK^F_\alpha$, as well as suitable boundary conditions at~$\ci^+_\pm$. In particular, the leading constraint equation on~$\ci^+$ is given by, 
\begin{equation}
\begin{split}\label{constrainteq1}
\d_{\b z} \lambda_{(+)} + \d_u \lambda_{(-)} = {e^2 \over 2} \b k_{(+)}~,
\end{split}
\end{equation}
which determines the~$u$-dependence of~$\lambda_{(-)}$ in terms of~$\lambda_{(+)}$ and~$k_{(+)}$. As in the bosonic case, we assume that the source~$k_{(+)}$ vanishes at~$\ci^+_\pm$. 

We would like to know how supersymmetry relates the fermionic boundary fields in~\eqref{photinolarger} and~\eqref{kappalarger} to the bosonic boundary fields in~\eqref{photonlarger} and~\eqref{currlarger}. Even though all four supercharges remain unbroken at null infinity, we will focus on supersymmetry transformations with constant spinor parameter~$\xi^{(-)}_\alpha$ and their complex conjugates, which are generated by the following supercharges
\begin{equation}\label{cqdef}
\CQ = \xi^{(-)\alpha} Q_\alpha~, \qquad \b \CQ = \b \xi^{(-)\alphadot} \b Q_\alphadot~, \qquad \left\{\CQ, \b \CQ\right\} = - 4 i \d_u~. 
\end{equation}
They are the position space analogues of the supercharges that act non-trivially on massless particle representations, as in~\eqref{susychimultstate}. The only non-vanishing commutators of~$\CQ$ with the boundary photon field~$A_\mu$ are given by
\begin{subequations}
\begin{align}
& \big[\CQ, A_{\b z} \big]  = \b \lambda_{(+)}~,\label{cqavar}  \\
&  \big[\CQ, A_r\big] = - \b \lambda_{(-)}~,
\end{align}
\end{subequations}
while the only non-vanishing anticommutators of~$\CQ$ with the boundary photino~$\lambda_{(\pm)}$ are
\begin{subequations}
\begin{align}
& \big\{\CQ, \lambda_{(+)} \big\} = 4 i \d_u A_z~,\\
& \big\{\CQ, \lambda_{(-)} \big\} = - D - 2i F_{ur} + 2 i F_{z \b z}~. \label{cqlamvar}
\end{align}
\end{subequations}
The action of~$\b \CQ$ on these fields can be obtained by taking the Hermitian conjugates of these formulas. Here~$F_{z \b z} = \d_z A_{\b z} - \d_{\b z} A_z$ is the large-$r$ limit of~$\CF_{z \b z}$, whose falloff is~$\CO(1)$, as discussed in~\eqref{flarger}. The auxiliary field~$\CD$ in the vector multiplet~\eqref{gaugemult} is a Lorentz scalar with IR scaling dimension~$\Delta_{\CD} = 2$, which according to~\eqref{rfalloff} falls off like~$\CD = {D \over r^2} + \CO(r^{-3})$. The fact that there are no residual powers of~$r$ in these formulas shows that the assumed large-$r$ falloffs are consistent with supersymmetry. 

It is straightforward to repeat the preceding discussion near~$\ci^-$. The photino fields on~$\ci^+$ and~$\ci^-$ must then be matched at spatial infinity. The appropriate matching conditions can be determined from the matching conditions~\eqref{Amatchcond} for the photon using supersymmetry. Combining the supersymmetry variation in~\eqref{cqavar} with the matching condition for~$A_z$ in~\eqref{Amatchcondi} leads to
\begin{equation}
\lambda_{(+)} \big|_{\ci^+_-} = \lambda^-_{(+)} \big |_{\ci^-_+}~.  \label{matchcond1}  
\end{equation}
Similarly, the supersymmetry variation in~\eqref{cqlamvar} and the matching condition for~$F_{ur}$ in~\eqref{Amatchcondii} imply that the~$u$-independent part of~$\lambda_{(-)}$ should be matched across spatial infinity. However, the constraint equation~\eqref{constrainteq1} implies that~$\lambda_{(-)} \big|_{\ci^+_-}$ does not exist, since 
\begin{equation}
\lambda_{(-)} \rightarrow u \left(-\d_{\b z} \lambda_{(+)} \big|_{\ci^+_-}\right) + \ell(z, \b z)  \qquad \text{as} \qquad u \rightarrow - \infty~.
\end{equation}
Instead, we should match the~$u$-independent term across spatial infinity, $\ell(z, \b z) = \ell^-(z, \b z)$, which can be expressed in terms of~$\lambda_{(-)}$ as follows,
\begin{equation}
\left( 1 - u \p_u \right) \lambda_{(-)}  \big|_{\ci^+_-} =   \left( 1 - u \p_u \right) \lambda^-_{(-)} \big|_{\ci^-_+}~.  \label{matchcond2}      
 \end{equation}

\subsection{Fermionic Asymptotic Symmetries} 

Consider the following fermionic charges on~$\ci^+$ and~$\ci^-$, for any complex-valued~$\chi(z, \b z)$,
\begin{equation}\label{fchidef}
\begin{split}
{\mathscr F}[\chi] &= \frac{1}{2e^2} \int d^2 z \, \chi(z, \b z)   \left( 1 - u \p_u \right) \lambda_{(-)}\big|_{\ci^+_-}~, \\
{\mathscr F}^-[\chi]  &=   \frac{1}{2e^2} \int d^2 z \, \chi(z, \b z)  \left( 1 - u \p_u \right) \lambda^-_{(-)} \big|_{\ci^-_+}~. 
\end{split}
\end{equation}
We can express them in a more covariant form by introducing a commuting, chiral spinor-valued function on~$S^2$, 
\begin{equation}\label{chiralchidef}
\chi_\alpha(z, \b z) = \chi(z, \b z) \xi_\alpha^{(+)}(z)~.
\end{equation}
Using the expansion~\eqref{lambdaexpansion} and the falloffs~\eqref{photinolarger}, we can write
\begin{equation}\label{covfchidef}
{\mathscr F}[\chi] = -\frac{1}{2e^2} \int d^2 z \, \chi^\alpha(z, \b z)   \left[\left( 1 - u \p_u \right) \left(\lim_{r\rightarrow \infty} r^2 \Lambda_\alpha(r, u, z, \b z)\right)\right] \Big|_{\ci^+_-}~,
\end{equation}
and similarly for~${\mathscr F}^-[\chi]$. Comparing the matching condition~\eqref{matchcond2} to~\eqref{fchidef} implies the conservation law\footnote{The matching condition \eqref{matchcond1} for $\lambda_{(+)}$ is similar to the discussion in section 3.3.4 of \cite{He:2020ifr} and does lead to additional constraints on the scattering amplitude, but we do not explore this here.}
\begin{equation}
\begin{split}\label{charge-eq}
{\mathscr F}[\chi]  = {\mathscr F}^-[\chi]~,
\end{split}
\end{equation}
and hence a Ward identity for the tree-level~$\CS$-matrix,
\begin{equation}\label{fermwardid}
{\mathscr F}[\chi] \CS - \CS {\mathscr F}^-[\chi] = 0~.
\end{equation}
In section~\ref{sec:sphotisjas} we will show that this identity gives rise to the positive-helicity soft photino theorem~\eqref{softphotino};~the Hermitian conjugate charges~${\mathscr F}^\dagger[\b \chi]$ lead to the negative-helicity case~\eqref{softphotinoneg}. In the remainder of this section we establish several basic properties of~${\mathscr F}[\chi]$.

Supersymmetry relates the fermionic symmetries~${\mathscr F}[\chi]$ defined in~\eqref{fchidef} to the bosonic asymptotic symmetries~${\mathscr E}[\ep]$ in~\eqref{bcharge}. For instance, we can use~\eqref{cqlamvar} to determine the anticommutators of the supercharges~$\CQ, \b \CQ$ that were singled out in~\eqref{cqdef} with~${\mathscr F}[\chi]$,\footnote{~Here we use the fact that~$D|_{\ci^+_\pm} = F_{z\b z}|_{\ci^+_\pm} = 0$. The first equation is obtained by solving for the auxiliary field~$\CD$ in~\eqref{gaugemult} in terms of the bosonic source~$\CK^B$ in~\eqref{jmult}, which is assumed to vanish at~$\ci^+_\pm$. The second equation follows from the fact that there are no magnetic charges.}
\begin{equation}\label{susyonFchi}
\big\{\CQ, {\mathscr F}[\chi]\big\} = -{i \over  e^2} \int d^2 z\, \chi(z, \b z) \, F_{ur} \big|_{\ci^+_-} = {i} {\mathscr E}[\chi]~, \qquad \big\{ \b \CQ, {\mathscr F}[\chi] \big \} = 0~.
\end{equation}
Note that the fermionic symmetry~${\mathscr F}[\chi]$, with complex parameter~$\chi(z, \b z)$, transforms into the bosonic symmetry~${\mathscr E}[\ep]$ with the same parameter, $\ep(z, \b z) = \chi(z, \b z)$. This shows that it is natural to allow complex~$\ep(z, \b z)$, as was discussed in~\cite{as,He:2015zea,Strominger:2015bla} and reviewed after~\eqref{bcomms}. More generally, we can use~\eqref{gaugesusy} and~\eqref{covfchidef} to express the commutator of an arbitrary supercharge with~${\mathscr F}[\chi]$ in the covariant form quoted in~\eqref{sm},
\begin{equation}\label{smii}
\big\{\zeta^\alpha Q_\alpha, {\mathscr F}[\chi]\big\} = i {\mathscr E}\left[\zeta^\alpha \chi_\alpha\right]~, \qquad \big\{\b Q_\alphadot, {\mathscr F}[\chi]\big\} = 0~.
\end{equation}
Here~$\zeta_\alpha$ is a commuting, constant spinor and~$\chi_\alpha(z, \b z)$ was defined in~\eqref{chiralchidef}. It can similarly be shown that the bosonic charges~${\mathscr E}[\ep]$ in~\eqref{ehard} are annihilated by all supercharges. This is expected from their interpretation as conventional gauge transformations that do not vanish at~$\ci^+$, since the latter commute with supersymmetry. 

Following the discussion of the bosonic case around~\eqref{bcharge}, we can express~${\mathscr F}[\chi]$ as an integral over~$\ci^+$ and use the constraint equation~\eqref{constrainteq1} to write it as a sum of hard and soft contributions,
\begin{equation}\label{fhardsoft}
{\mathscr F}[\chi] = {\mathscr F}^\text{h}[\chi] + {\mathscr F}^s[\chi]~.
\end{equation}
The hard charge is given by
\begin{equation}
\label{Fchidef1}
{\mathscr F}^\text{h}[\chi] = \frac{1}{4} \int  du d^2 z \, \chi(z, \b z)  u \p _u \b k_{(+)}  +  \frac{1}{2e^2} \int d^2 z \, \chi(z, \b z)   \left( 1 - u \p_u \right) \lambda_{(-)} \big|_{\ci^+_+}~. 
\end{equation}
As in~\eqref{ehard}, the first term represents the contribution of massless charged particles that couple to the photino, while the second term is nontrivial only if there are  massive charged particles passing through future timelike infinity. Since we are considering theories without massive particles, we set~$ \left( 1 - u \p_u \right) \lambda_{(-)} \big|_{\ci^+_+} = 0$.\footnote{~Following~\cite{Campiglia:2015qka,Kapec:2015ena}, it should be possible to incorporate massive particles by appropriately taking into account their semiclassical photino field as they pass through timelike infinity.} The supersymmetry transformation~\eqref{cqlamvar} turns this condition into~$F_{ur} \big|_{\ci^+_+} = 0$, which was imposed after~\eqref{ehard}. We can compute the following anticommutators with the supercharges singled out in~\eqref{cqdef},
\begin{equation}\label{qoffhard}
\big\{\CQ, {\mathscr F}^\text{h}[\chi]\big\} = {i \over 2} \int du d^2 z \, \chi(z, \b z) j_u = i {\mathscr E}^\text{h}[\chi]~, \qquad \big\{\b \CQ,  {\mathscr F}^\text{h}[\chi]\big\} = 0~,
\end{equation}
up to boundary terms at~$\ci^+_\pm$ that involve the sources and hence vanish by assumption. In section~\ref{sec:sphotisjas} we will use these relations to determine the action of the hard charges~${\mathscr F}^\text{h}[\chi]$ on asymptotic scattering states.

The soft charges in~\eqref{fhardsoft} are given by\begin{equation}\label{fsoftdef}
{\mathscr F}^\text{s}[\chi] = \frac{1}{2 \pi} \int d^2 z  \,  \p_\bz \chi(z, \b z) \, \omega^\text{s}~, \qquad \omega^\text{s}  = {\pi \over e^2} \int du \, u\d_u \lambda_{(+)}~.
\end{equation}
Here we have defined a soft photino current~$\omega^\text{s}$.\footnote{~Note that the operator~$\int du \, \d_u \lambda_{(+)}$, which is similar to soft photon current~$j_z^\text{s}$ defined in~\eqref{esoft}, can be shown to vanish inside~$\CS$-matrix elements by expressing it in terms of creation and annihilation operators (see appendix~\ref{modeexp}) and comparing to the soft photino theorem~\eqref{softphotino}.} Under a Lorentz transformation~\eqref{lordef}, it changes as follows,
\begin{equation}\label{omegalort}
\delta_Y \omega^\text{s} = \left(\half \d_z Y^z + Y^z \d_z + Y^{\b z} \d_{\b z}\right) \omega^\text{s}~,
\end{equation}
up to boundary terms that vanish as long as~$\lambda_{(+)}$ asymptotes to a~$u$-independent function of~$z, \b z$ sufficiently rapidly at~$\ci^+_\pm$.\footnote{~It is sufficient to assume that~$\lambda_{(+)} = \lambda_{(+)}\big|_{\ci^+_\pm} + \CO\big(|u|^{-(1+\delta)}\big)$, with~$\delta > 0$, as~$u \rightarrow \pm \infty$.} The Lorentz transformation~\eqref{omegalort} shows that the soft photino current~$\omega^\text{s}$ is a two-dimensional field with~$SL(2, \C)$ conformal weights~$h = \half$ and $\b h = 0$, i.e.~it is a left-moving spin-$\half$ current. Under the supercharges in~\eqref{cqdef}, the soft photino current~$\omega^\text{s}$ transforms into the soft photon current~$j_z^\text{s}$ defined in~\eqref{esoft} as follows,
\begin{equation}
\big\{\CQ, \omega^\text{s}\big\} =i j_z^\text{s}~, \qquad \big\{\b \CQ, \omega^\text{s}\big\} = 0~.
\end{equation}

In order to understand the action of the soft charges~\eqref{fsoftdef} on the photino, it is convenient to rewrite them as follows,\footnote{~Given a function~$f(u)$ such that~$\displaystyle \lim_{u \rightarrow \pm \infty} f(u)$ exists, but is nonzero, we have the following identity,
\begin{equation}
\int_{-\infty}^\infty du \, u f'(u) = - \lim_{\omega \rightarrow 0}  \int_{-\infty}^\infty du \, \cos(\omega u) f(u)~,
\end{equation}
which amounts to integrating by parts but dropping the divergent boundary terms.}
\begin{equation}
{\mathscr F}^\text{s}[\chi] = -{1 \over 2 e^2} \lim_{\omega \rightarrow 0} \int du d^2 z \, \d_{\b z} \chi(z, \b z) \, \cos(\omega u) \lambda_{(+)}~.
\end{equation}
In terms of creation and annihilation operators (see appendix~\ref{modeexp}),
\begin{equation}\label{fsoftccdag}
{\mathscr F}^\text{s}[\chi] = {i \over 4 \sqrt 2 e\pi} \lim_{\omega \rightarrow 0} \sqrt \omega \int d^2 z \, \d_{\b z} \chi(z, \b z) \, \left(a_{\b \Lambda, +}(\omega, z, \b z) - a_{\Lambda, -}^\dagger(\omega, z, \b z)\right)~.
\end{equation}
This shows that~${\mathscr F}^\text{s}[\chi]$ acts on zero-momentum photini. Using this expression, as well as the mode expansion for~$\lambda_{(+)}$ and the anticommutation relations for creation and annihilation operators in appendix~\ref{modeexp}, it can be checked that
\begin{equation}\label{fsoftcomm}
\big\{{\mathscr F}^\text{s}[\chi], \lambda_{(+)}(u, z, \b z) \big\} = 0~, \qquad \big\{{\mathscr F}^\text{s}[\chi], \b \lambda_{(+)} (u, z, \b z)\big\} = - \d_{\b z} \chi(z, \b z)~.
\end{equation}
Thus, $\b \lambda_{(+)}$ shifts inhomogeneously whenever~$\d_{\b z} \chi(z, \b z) \neq 0$. Just as in the bosonic case~\eqref{bcomms}, we interpret this as spontaneous breaking of the corresponding charges~${\mathscr F}^\text{s}[\chi]$. The~$u$-independent part of~$\b \lambda_{(+)}$ furnishes the corresponding Goldstone fermions. Similar comments apply to~${\mathscr F}^{\dagger}[\chi]$, which shifts~$\lambda_{(+)}$ by~$- \d_z \b \chi(z, \b z)$.

\section{Soft Photino Theorem from Asymptotic Fermionic Symmetries} 
\label{sec:sphotisjas}

\subsection{Fermionic Ward Identity for Scattering Amplitudes}
\label{sec:fwidampl}

In the previous section we argued for the existence of a fermionic asymptotic symmetry~${\mathscr F}[\chi]$, which is classically conserved (see~\eqref{charge-eq}) and hence leads to a Ward identity~\eqref{fermwardid} for the tree-level~$\CS$-matrix, 
\begin{equation}\label{fermwardii}
{\mathscr F}[\chi] \CS - \CS {\mathscr F}^-[\chi] = 0~.
\end{equation}
We will now show that this Ward identity is nothing but the soft photino theorem for the case of an outgoing positive-helicity photino (equivalently, by crossing symmetry, an ingoing negative helicity photino), which we repeat for convenience, 
\begin{equation}\label{softphotinoii}
\CM^{\text{out},+}_{n+1} ~~~ \longrightarrow ~~~ { \sqrt 2 i e} \sum_{i = m+1}^n \,  {q_i \over \eta(p_{n+1}) \eta(p_i)} \left({\mathscr F}_i \CM_n\right) - { \sqrt 2 i e} \sum_{i = 1}^m \, {q_i \over \eta(p_{n+1}) \eta(p_i)} \left({\mathscr F}_i \CM_n\right)~. 
\end{equation}
Here~$p_{n+1} \rightarrow 0$ is the momentum of the soft photino. Analogously, the Ward identity for~${\mathscr F}^\dagger[\b \chi]$ leads to the soft photino theorem~\eqref{softphotinoneg} for an outgoing negative helicity photino. 

We begin by translating~\eqref{softphotinoii} from momentum to position space. As explained in section~\ref{sec:prelim}, we can express the null momenta~$p_i$ of the~$n+1$ external particles in terms of variables~$\omega_i, z_i, \b z_i$, using the parametrization in~\eqref{pdef}. In particular, the spinor-helicity variables corresponding to the~$p_i$ are given by~\eqref{etaxirel}, so that
\begin{equation}
\eta^\alpha(p_{n+1}) \eta_{\alpha}(p_{i}) = 2 \sqrt{\omega_{n+1} \omega_i} \left(z_{n+1} - z_i\right)~, \qquad (i = 1, \ldots, n)~.
\end{equation} 
In this parametrization, the soft photino theorem can be written as follows,
\begin{equation}\label{softphotinopos}
\sqrt {2 \omega_{n+1}} \CM^{\text{out},+}_{n+1}~ \longrightarrow~{ i e} \sum_{i = m+1}^n \,  {q_i \over \sqrt{\omega_i}} {1 \over  z_{n+1} - z_i} \left({\mathscr F}_i \CM_n\right) - { i e} \sum_{i = 1}^m \, {q_i \over \sqrt{\omega_i}} {1 \over  z_{n+1} - z_i}  \left({\mathscr F}_i \CM_n\right)~. 
\end{equation}

In order to reproduce this result, we take the matrix element of the Ward identity~\eqref{fermwardii} between an~$m$-particle in-state~$\ket{1 \, ; \, \ldots \, ; \, m}$ and an~$(n-m)$-particle out-state~$\bra{m+1 \, ; \, \ldots \, ; \, n}$. All in- and outgoing particles (some of which could be photini) are hard, i.e.~they have non-vanishing momenta. Writing~${\mathscr F}[\chi] = {\mathscr F}^\text{h}[\chi] + {\mathscr F}^\text{s}[\chi]$ as a sum of hard and soft contributions, as in~\eqref{fhardsoft}, and similarly for~${\mathscr F}^-[\chi]$, we obtain
\begin{equation}
\begin{split}\label{wardid1}
 \bra{m+1 \, ; \, \ldots \, ; \, n}   \, {\mathscr F}^\text{s}[\chi] \CS & - \CS {\mathscr F}^{\text{s}-}[\chi] \,   \ket{1 \, ; \, \ldots \, ; \, m} = \\
& -  \bra{m+1 \, ; \, \ldots \, ; \, n}   \, {\mathscr F}^\text{h}[\chi] \CS - \CS {\mathscr F}^{\text{h}-}[\chi] \,   \ket{1 \, ; \, \ldots \, ; \, m} ~. 
\end{split}
\end{equation}
To proceed, we need to know the action of the soft and hard charges on asymptotic scattering states. The soft charge was expressed in terms of photino creation and annihilation operators in~\eqref{fsoftccdag}. It creates an outgoing positive-helicity photino and an ingoing negative-helicity photino of zero momentum. Crossing symmetry implies that these two contributions lead to identical~$\CS$-matrix elements, so that we can write the left-hand side of~\eqref{wardid1} as the~$\omega_{n+1} \rightarrow 0$ limit of
\begin{equation}\label{lhs}
{ i \sqrt{\omega_{n+1}} \over 2 \sqrt 2 e \pi} \int d^2w \, \d_{\b w} \chi(w, \b w) \, \bra{m+1 \, ; \, \ldots \, ; \, n\, ; \, \b \Lambda, p(\omega_{n+1}, w, \b w)}    \CS   \big|1 \, ; \, \ldots \, ; \, m\big\rangle~.
\end{equation}
The action of the hard charges on asymptotic states will be derived section~\ref{sec:hardchmat} below, where it is shown that
\begin{equation}\label{fhardmat}
\begin{split}
& {\mathscr F}^\text{h}[\chi] \big|f, p(\omega, z, \b z), s \big\rangle = -{q_f \over 2 \sqrt \omega} \chi(z, \b z) {\mathscr F} \big|f, p(\omega, z, \b z), s \big\rangle~,\\
& {\mathscr F}^{\text{h} \dagger} [\b \chi] \big|f, p(\omega, z, \b z), s \big\rangle = -{q_f \over 2 \sqrt \omega} \b \chi(z, \b z) {\mathscr F}^\dagger \big|f, p(\omega, z, \b z), s \big\rangle~.
\end{split}
\end{equation}
Here~$q_f$ is the electric charge of the state labeled by~$f \in \{\Phi, \b \Phi, \Psi, \b \Psi\}$. The operator~${\mathscr F}$ and its Hermitian conjugate~$\mathscr F^\dagger$ appear in the soft theorem~\eqref{softphotinoii}. Its action on chiral and anti-chiral matter states was defined in~\eqref{scriptFdef}. 

If we choose~$\chi(w, \b w) = {1 \over z_{n+1} - w}$, where~$z_{n+1}$ is the~$z$-value parametrizing the momentum~$p_{n+1}$ of the soft photino, the Ward identity collapses to the soft theorem~\eqref{softphotinoii}. As in the bosonic case~\cite{hmps}, the argument can be reversed to deduce the Ward identity -- and hence the underlying symmetries  -- from the soft theorem, which establishes their equivalence.

\subsection{Action of the Fermionic Charges on Matter Fields}
\label{sec:hardchmat}

Here we show that the action of the hard fermionic charges~${\mathscr F}^\text{h}[\chi]$ on asymptotic states is given by~\eqref{fhardmat}, thereby completing the argument of section~\ref{sec:fwidampl}. We will do this by using the supersymmetry relations~\eqref{qoffhard}, 
\begin{equation}\label{qoffhardii}
\big\{\CQ, {\mathscr F}^\text{h}[\chi]\big\} = i {\mathscr E}^\text{h}[\chi]~, \qquad \big\{\b \CQ, {\mathscr F}^\text{h}[\chi]\big\} = 0~.
\end{equation}
Here~${\mathscr E}^\text{h}[\chi]$ are the hard bosonic charges, whose action~\eqref{bcomms} on boundary fields~$f_q(u, z, \b z)$ of electric charge~$q$ is given by
\begin{equation}\label{bcommsii}
\big[{\mathscr E}^\text{h}[\ep], f_q(u, z, \b z)\big] = - q \ep(z, \b z) f_q(u, z, \b z)~.
\end{equation}
Given the action of the supercharges~$\CQ, \b \CQ$ on charged boundary fields, we can extract the action of~${\mathscr F}^\text{h}[\chi]$ on such fields from~\eqref{qoffhardii} and~\eqref{bcommsii}. The same logic was used in section~\ref{sec:softphotinothm} to relate the matrix elements of the fermionic source~$\b \CK^F_\alphadot$ to those of the electric current~$\CJ_\mu$. 

For our present purposes, all charged fields reside in massless chiral or anti-chiral multiplets. A chiral multiplet consists of a complex scalar~$\Phi$, whose asymptotic expansion near~$\ci^+$ was described in~\eqref{scalarlarger},
\begin{equation}\label{scalarlargerii}
\Phi(u, r, z, \b z) = {1 \over r} \phi(u, z, \b z) + \CO(r^{-2})~,
\end{equation}
and a left-handed fermion~$\Psi_\alpha$, whose large-$r$ behavior is identical to that of the photino~$\Lambda_\alpha$, which was discussed around~\eqref{lambdaexpansion} and~\eqref{photinolarger},
\begin{equation}\label{psilarger}
\Psi_\alpha(u, r, z, \b z) = {1 \over r} \psi(u, z, \b z) \xi_\alpha^{(+)} + \CO(r^{-2})~.
\end{equation}
If the chiral multiplet has charge~$q$, then so do the boundary fields~$\phi$ and~$\psi$, i.e.
\begin{equation}\label{eonphipsi}
\big[{\mathscr E}^\text{h}[\ep], \phi(u, z, \b z) \big] = - q \ep(z, \b z) \phi(u, z, \b z)~, \qquad \big[{\mathscr E}^\text{h}[\ep], \psi(u, z, \b z) \big] = - q \ep(z, \b z) \psi(u, z, \b z)~. 
\end{equation}
Given the asymptotic expansions~\eqref{scalarlargerii} and~\eqref{psilarger}, we obtain the following transformation rules for the boundary fields~$\phi, \psi$ under the supercharges~$\CQ, \b \CQ$ singled out in~\eqref{cqdef},\footnote{~In principle the anticommutator~$\{\CQ, \psi\}$, which falls off as~$\CO(r^{-1})$ at large~$r$, could receive a contribution from the dimension~$2$ scalar auxiliary field~$F$ in the chiral multiplet, but according to~\eqref{rfalloff} such a field falls off as~$\CO(r^{-2})$ and hence it does not contribute.}
\begin{subequations}\label{qonchiral}
\begin{align}
& \left[ \CQ , \phi \right] = \sqrt{2} i \psi~, & &  \left[ {\bar \CQ} , \phi(u, z, \b z) \right] = 0~, \label{phisusy}\\
& \left\{\CQ, \psi(u, z, \b z)  \right\} = 0  ~,& &  \left\{ {\bar \CQ}  , \psi(u, z, \b z) \right\}  = - 2 \sqrt{2} \p_u \phi(u, z, \b z)~. \label{psisusy}
\end{align}
\end{subequations}

Given the transformation properties~\eqref{eonphipsi} and~\eqref{qonchiral} of the chiral multiplet fields under the bosonic symmetry~${\mathscr E}^\text{h}[\ep]$ and the supersymmetries~$\CQ, \b \CQ$, the commutators in~\eqref{qoffhardii} are only consistent if the fermionic symmetry~${\mathscr F}^\text{h}[\chi]$ acts as follows,
\begin{equation}
\begin{split}\label{Fchi1}
\left[{\mathscr F}^\text{h}[\chi], \phi(u, z, \b z) \right] = 0~, \qquad \left\{ {\mathscr F}^\text{h}[\chi] , \psi(u, z, \b z)  \right\} = -  \frac{q}{\sqrt{2}} \chi(z, \b z)   \phi(u , z, \b z)~. 
\end{split}
\end{equation}
The first commutator can be understood as a consequence of the~$U(1)_R$ symmetry that is expected to emerge at the superconformal IR fixed point that governs the dynamics near null infinity. Since~${\mathscr F}[\chi]$ is linear in the photino (see~\eqref{fchidef}), it has~$R$-charge~$+1$. (We take the~$R$-charge of~$Q_\alpha$ to be~$-1$.) The electric and~$U(1)_R$ charges of the first commutator in~\eqref{Fchi1} are not consistent with any fermionic field in the chiral multiplet, and hence it must vanish. 

For the anti-chiral multiplet  of charge~$-q$, which is described by taking the Hermitian conjugates of~\eqref{scalarlargerii}, \eqref{psilarger}, \eqref{eonphipsi}, and~\eqref{qonchiral}, the consistency of~\eqref{qoffhardii} requires
\begin{subequations}
\begin{align}
& \left\{{\mathscr F}^\text{h}[\chi], \b \psi(u, z, \b z)\right\} = 0~,\label{fonpsib}\\
& \left[{\mathscr F}^\text{h}[\chi], \d_u \b \phi(u, z, \b z)\right] = -{i q \over 2 \sqrt 2} \chi(z, \b z) \b \psi(u, z, \b z)~,\label{fonduphib} \\
& \left\{\CQ, \left[{\mathscr F}^\text{h}[\chi], \b \phi(u, z, \b z)\right]\right\} = i q \chi(z, \b z) \b \phi(u, z, \b z)~. \label{qfbphi}
\end{align}
\end{subequations}
As above, the first equation~\eqref{fonpsib} is due to the electric and~$U(1)_R$ charges of the fields. While~\eqref{fonduphib} shows that~$\d_u \b \phi$ has a local transformation rule, it follows from~\eqref{qfbphi} that this does not lead to a local transformation rule for~$\b \phi$ itself. If it did, then~$\b \phi$ would be~$\CQ$-exact, which is not the case because~$\b \phi$ is the bottom component of the supermultiplet in~\eqref{qonchiral}.

The (anti-) commutators in~\eqref{Fchi1}, \eqref{fonpsib}, and~\eqref{fonduphib} are sufficient to establish the action of~${\mathscr F}^\text{h}[\chi]$ and~${\mathscr F}^{\text{h} \dagger}[\b \chi]$ on asymptotic states. Using the mode expansions in appendix~\ref{modeexp} and the fact that~${\mathscr F}^\text{h}[\chi]$ annihilates the vacuum,\footnote{~Recall that the soft charges~${\mathscr F}^\text{s}[\chi]$ are spontaneously broken, since they shift the photino as in~\eqref{fsoftcomm} and hence do not annihilate the vacuum. However, this is not the case for the hard charges~${\mathscr F}^\text{h}[\chi]$.} we find that
\begin{equation}\label{fhonstatesfin}
\begin{split}
& {\mathscr F}^\text{h}[\chi] \, \big|\Phi, p(\omega, z, \b z)\big\rangle = {\mathscr F}^\text{h}[\chi] \, \big |\b \Psi, p(\omega, z, \b z), +\big \rangle = 0~, \\
& {\mathscr F}^\text{h}[\chi] \, \big|\Psi, p(\omega, z, \b z), -\big\rangle = - {q \over 2 \sqrt \omega} \chi(z, \b z) \big| \Phi, p(\omega, z, \b z)\big\rangle~,\\
& {\mathscr F}^\text{h}[\chi] \, \big|\b \Phi, p(\omega, z, \b z)\big\rangle = -{q \over 2 \sqrt \omega} \chi(z, \b z) \big| \b \Psi, p(\omega, z, \b z), +\big\rangle~.
\end{split}
\end{equation} 
Here the null momenta of the asymptotic  states are parametrized in terms of~$\omega, z, \b z$ as in~\eqref{pdef}. We can express~\eqref{fhonstatesfin} in terms of the operator~$\mathscr F$, whose action on asymptotic states was defined in~\eqref{scriptFdef},
\begin{equation}
{\mathscr F} |\Phi, p\rangle = {\mathscr F} |\b \Psi, p, +\rangle = 0~, \qquad {\mathscr F}|\b \Phi, p\rangle = |\b \Psi, p, +\rangle~, \qquad  {\mathscr F}|\Psi, p, -\rangle = - |\Phi, p\rangle~.
\end{equation}
Since~$|\b \Phi, p\rangle$ has charge~$q$ and~$|\Psi, p, -\rangle$ has charge~$-q$, we can express~\eqref{fhonstatesfin} as follows,
\begin{equation}
\begin{split}\label{Fchione-particle}
{\mathscr F}^\text{h}[\chi]    \ket{ f , p , s } = - \frac{q_f }{ 2 \sqrt{ \omega } }\chi(z,\bz) {\mathscr F} \ket{ f , p , s }~,
\end{split}
\end{equation}
where~$q_f$ is the electric charge of the state. It is straightforward to repeat the preceding discussion for the Hermitian conjugate charges. They obey
\begin{equation}
\begin{split}\label{fdaggeronepart}
{\mathscr F}^{\text{h}\dagger}[\b \chi]    \ket{ f , p , s } = - \frac{q_f }{ 2 \sqrt{ \omega } }\b \chi(z,\bz) {\mathscr F}^\dagger \ket{ f , p , s }~,
\end{split}
\end{equation}
where the action of~$\mathscr F^\dagger$ on one-particle asymptotic states was defined in~\eqref{scriptFdef}. Together with~\eqref{Fchione-particle}, this establishes the relations stated in~\eqref{fhardmat}.

\section*{Acknowledgements}
 
\noindent We are grateful to H.~Elvang, D.~Kapec, V.~Lysov, S.~Naculich, S.~Pasterski, and B.~Schwab for discussions. This work is supported by the Fundamental Laws Initiative at Harvard University, as well as DOE grant DE-SC0007870 and NSF grant PHY-1067976.

\appendix

\section{Conventions}\label{app:conventions}

Unless stated otherwise we follow the conventions of~\cite{Wess:1992cp}. Here we collect various useful formulas related to the choice of coordinates and spinor basis introduced in section~\ref{sec:prelim}. 

\subsection{Coordinates} 

Flat Minkowski coordinates are denoted by~$y^a$. We work in curvilinear coordinates $x^\mu = (u, r, z, \b z)$, which are related to the~$y^a$ as follows,
\begin{equation}\label{yzcoordsapp}
\begin{split}
y^0 &= \frac{1}{2} \left(u + r  \left( 1+|z|^2\right) \right)~, \\
y^1 &= \frac{r}{2} \left( z + \bz \right)~, \\
y^2 &= - \frac{ir}{2} \left( z - \bz \right)~, \\
y^3 &= - \frac{1}{2} \left( u - r  \left( 1 - |z|^2 \right)  \right)~. 
\end{split}
\end{equation}
The Minkowski metric is given by 
\begin{equation}
\begin{split}
ds^2 = \eta_{ab} dy^a dy^b = - du dr + r^2 dz d\bz~.
\end{split}
\end{equation}
The~$\delta$-function in the~$z, \b z$ plane is normalized as follows,  
\begin{equation}
\begin{split}
g(w,\b w)  = \int d^2 z \, \delta^{(2)} (z-w) g(z,\b z) ~, \qquad  d^2 z = i dz \wedge d \bz~.
\end{split}
\end{equation}

In this paper, we are interested in the regions~$r \rightarrow \pm \infty$, with~$u, z, \b z$ finite. When $r \rightarrow +\infty$, the coordinate charge in~\eqref{yzcoordsapp} asymptotes to 
\begin{equation}\label{scripluscoords}
\begin{split}
y^0 - |\vec{y}\,| \rightarrow { u \over 1 + |z|^2}~, \qquad y^0 + |\vec{y}\,| \rightarrow \left(1+|z|^2\right) r~, \qquad w\left(\vec y\,\right) \rightarrow z~.
\end{split}
\end{equation}
Here~$w\left(\vec y\,\right) \in \C$ is the standard stereographic coordinate on the unit~$S^2$ defined by~${\vec{y} \over |\vec{y}\,|} \in \R^3$. This shows that the large-$r$ region parametrized by~$u, z, \b z$ is precisely~$\ci^+$, with the exception of the point~$z = \infty$. Similarly, the~$r \rightarrow -\infty$ limit of~\eqref{yzcoordsapp} is given by
\begin{equation}\label{scriminuscoords}
\begin{split}
y^0 - |\vec{y}\,| \rightarrow \left(1+|z|^2\right) r ~, \qquad y^0 + |\vec{y}\,| \rightarrow { u \over 1 + |z|^2}~, \qquad w\left(\vec y\,\right) \rightarrow -{1 \over \b z}~,
\end{split}
\end{equation}
which correctly describes~$\ci^-$ (again with the exception of a point on~$S^2$). Note that the same~$z$-coordinate in~\eqref{scripluscoords} and~\eqref{scriminuscoords} leads to~$w$-coordinates that are related by the antipodal map~$w \rightarrow -{1 \over \b w}$.

\subsection{Spinors} 

In order to discuss spinors in the curvilinear coordinates~$x^\mu$, we choose the vielbein\begin{equation}
\begin{split}\label{vielbeinapp}
e^a_\mu = \pd{y^a}{x^\mu}~,
\end{split}
\end{equation}
with~$y^a(x^\mu)$ as in~\eqref{yzcoordsapp}, which leads to a vanishing spin connection. In this frame, we employ the following bases for left- and right-handed spinors, 
\begin{equation}
\begin{split}\label{xidefapp}
\xi^{(+)}_\a(z) =  \left( \begin{array}{c} 1 \\ z \end{array} \right) ~, \qquad \xi^{(-)}_\a =  \left( \begin{array}{c} 0 \\ 1 \end{array} \right)~, \qquad \b \xi^{(+)}_\da(\b z) =  \left( \begin{array}{c} 1 \\ \b z \end{array} \right) ~, \qquad \b \xi^{(-)}_\da =  \left( \begin{array}{c} 0 \\ 1 \end{array} \right)~. 
\end{split}
\end{equation}
The~$\sigma$-matrices can then be expressed as simple bilinears,
\eqref{xidefapp},
\begin{equation}
\begin{split}
\sigma^u_{\a\alphadot} &= - 2 \xi^{(+)}_\a {\bar \xi}^{(+)}_\alphadot~, \quad 
 \sigma^r_{\a\alphadot} = - 2 \xi^{(-)}_\a {\bar \xi}^{(-)}_\alphadot~, \quad
  \sigma^z_{\a\alphadot} = \frac{2}{r} \xi^{(+)}_\a {\bar \xi}^{(-)}_\alphadot~, \quad
   \sigma^\bz_{\a\alphadot} = \frac{2}{r}  \xi^{(-)}_\a {\bar \xi}^{(+)}_\alphadot~. 
\end{split}
\end{equation}
Similar formulas for the~$\b \sigma$-matrices can be obtained by raising the spinor indices. The left-handed Lorentz generators~$\sigma_{\mu\nu} = {1 \over 4} \left(\sigma_\mu \b \sigma_\nu - \sigma_\nu \b \sigma_\mu\right)$ can be expressed as
\begin{equation}
\begin{split}
( \sigma_{ur} )_{\a\beta}  &= - \frac{1}{4} \left( \xi^{(+)}_\a \xi^{(-)}_\beta + \xi^{(-)}_\a \xi^{(+)}_\beta \right) ~, \\
(\sigma_{uz} )_{\a\beta} &= - \frac{r}{2} \, \xi^{(-)}_\a \xi^{(-)}_\beta ~, \\
(\sigma_{r\bz} )_{\a\beta} &=  \frac{r}{2} \,  \xi^{(+)}_\a \xi^{(+)}_\beta ~, \\
( \sigma_{z\bz} )_{\a\beta} &= \frac{r^2}{4} \left( \xi^{(+)}_\a \xi^{(-)}_\beta + \xi^{(-)}_\a \xi^{(+)}_\beta \right) ~, \\
( \sigma_{u\bz} )_{\a\beta} &= ( \sigma_{rz} )_{\a\beta} = 0 ~. 
\end{split}
\end{equation}
The right-handed Lorentz generators are given by Hermitian conjugation, since~$\b \sigma_{\mu\nu} = - \left(\sigma_{\mu\nu}\right)^\dagger$. 

The Lie derivatives of left- and right-handed spinor fields~$\Psi_\alpha$ and~$\b \Psi_\alphadot$ along a Killing vector~$K^\mu$ are given by 
\begin{equation}
\begin{split}
\CL_K \Psi_\a &= K^\mu \grad_\mu \Psi_\a - \frac{1}{2} \nabla_\mu K_\nu ( \sigma^{\mu\nu} )_\a{}^\beta \Psi_\beta~, \\
\CL_K {\bar \Psi}^\da &= K^\mu \grad_\mu {\bar \Psi}^\da - \frac{1}{2} \nabla_\mu K_\nu  ( {\bar \sigma}^{\mu\nu} )^\da{}_\db \b \Psi^\betadot~.
\end{split}
\end{equation}
Since the spin connection vanishes in our frame~\eqref{vielbeinapp}, the covariant derivatives of the spinors reduce to ordinary partial derivatives.

\subsection{Null Momenta} 

In flat Minkowski coordinates, we can parametrize null momenta~$p^a$ as follows, 
\begin{align}\label{momparameter}
p^a = \w \, \left(1+|z|^2, z+\bz, -i(z-\bz),1-|z|^2\right)~.
\end{align}
Here~$\omega$ has dimension of mass, but it is only equal to the energy~$p^0$ when~$z = 0$. In this parametrization, the Lorentz-invariant~$\delta$-function in momentum space is given by
\begin{equation}
\begin{split}
(2p^0) \delta^{(3)} \,  \big( \vec{p} - \vec{p}\,'  \big)  = \frac{1}{\omega}  \delta \left( \omega - \omega' \right) \delta^{(2)} \left( z - z' \right)~. 
\end{split}
\end{equation}

Since~$\det\, ( p \cdot \sigma ) =- p^2 =  0$, we can express the null momentum as a product of spinor-helicity variables,
\begin{equation}\label{shvdefapp}
p_\mu \sigma^\mu_{\alpha\alphadot} = \eta_\alpha(p) \b \eta_\alphadot(p)~.
\end{equation}
In terms of the usual angle and square brackets, $\eta_\alpha(p_i) = i\rangle_\alpha$ and~$\b \eta^\alphadot(p_i) =  i ]^\alphadot$. If we parametrize the null momentum~$p^a$ as in~\eqref{momparameter}, the spinor-helicity variables are related to the basis spinors in~\eqref{xidefapp} as follows,  
\begin{equation}\label{etaxirelapp}
\eta_\alpha(p) = \sqrt{2 \omega} \xi^{(+)}_\alpha(z)~, \qquad \b \eta_\alphadot(p) = \sqrt{2 \omega} \, \b \xi^{(+)}_\alphadot(\b z)~. 
\end{equation}
This fixes our convention for their little group phases, which is not determined by~\eqref{shvdefapp}.

\section{Mode Expansions}
\label{modeexp}

The free-field mode expansions that appear in this paper are given by 
\begin{equation}
\begin{split}\label{modexp}
\A_{\mu} (x) &= e \sum_{ \pm} \int \frac{d^3p}{(2\pi)^3} \frac{1}{2 p^0 }  \left[\left( \ve^{(\pm)}_{\mu} (p)\right)^{*} \, a_{\CF, \pm}  (p) e^{i p \cdot x } +   \ve^{(\pm)}_{\mu} (p) \, a^{\dagger}_{\CF, \pm}  (p) e^{- i p \cdot x }   \right] ~,  \\
\Lambda_\a (x) &= e \int \frac{d^3p}{(2\pi)^3} \frac{\eta_\a( {p}) }{2 p^0 }  \left[ a_{\b \Lambda, +} ( {p}) e^{i p \cdot x } + a_{\Lambda, -}^{\dagger}( {p}) e^{- i p \cdot x }   \right] ~, \\
\Phi(x) &=  \int \frac{d^3p}{(2\pi)^3} \frac{1 }{2 p^0 }  \left[ a_{\b \Phi} ( {p}) e^{i p \cdot x } + { a}^{\dagger}_{\Phi} ( {p}) e^{- i p \cdot x }   \right] ~, \\
\Psi_{\a}(x) &=  \int \frac{d^3p}{(2\pi)^3} \frac{\eta_{\a}(p) }{2 p^0 }  \left[ a_{\b \Psi, +} ( {p}) e^{i p \cdot x } + a_{\Psi, -}^{\dagger}( {p}) e^{- i p \cdot x }   \right] ~. 
\end{split}
\end{equation}
Here the photon and photino fields have an extra factor of the gauge coupling~$e$, because their kinetic terms in~\eqref{action} are not canonically normalized. The non-vanishing (anti-) commutation relations satisfied by the creation and annihilation operators are given by
\begin{equation}
\begin{split}
& [ a_{\CF, \pm} (p) , a_{\CF, \pm}^{\dagger} (p')  ] = [ a_\Phi(p) , a^\dagger_\Phi(p')] =  [ a_{\b \Phi}(p) , a_{\b \Phi}^\dagger(p')] =   \left( 2\pi \right)^3 \, (2 p^0) \, \delta^3 \big( \vec{p} - \vec{p}\,' \big) ~, \\
& \{ a_{\Lambda, -} (p) , a_{\Lambda, -}^{\dagger} (p') \} = \{ a_{\b \Lambda, +} (p) , a_{\b \Lambda, +}^{\dagger} (p') \} \\
& \qquad \qquad = \{ a_{\Psi, -} (p) , a_{\Psi, -}^{\dagger} (p') \} = \{ a_{\b \Psi, +} (p) , a_{\b \Psi, +}^{\dagger} (p') \} = \left( 2\pi \right)^3  \, (2 p^0) \, \delta^3 \big( \vec{p} - \vec{p}\,' \big)~. 
\end{split}
\end{equation}
The one-particle states that are used throughout the paper are given by \begin{align}
& \ket{ \F , p , \pm } = a_{\CF, \pm}^{\dagger} (p)  \ket{0}~, & & \cr
& \ket{ \Lambda , p, - } = a_{\Lambda, -}^{\dagger}( {p}) \ket{0}~, & &\ket{ {\bar \Lambda} , p, + } = a_{\b \Lambda, +}^{\dagger}( {p}) \ket{0}~,  \cr
& \ket{ \Phi , p } =  {a}_\Phi^{\dagger}( {p}) \ket{0}~, & & \ket{ {\bar \Phi} , p } =  a^{\dagger}_{\b \Phi}( {p}) \ket{0}  ~, \cr
& \ket{ \Psi, p, - }  = a_{\Psi, -}^{\dagger}( {p}) \ket{0} , & & \ket{ {\bar \Psi} , p, + } = a_{\b \Psi, +}^{\dagger}( {p}) \ket{0}~.
\end{align} 

The mode expansions for the boundary fields $A_{z}, \lambda_{(+)}, \phi, \psi$ are obtained by taking a suitable large-$r$ limit of the formulas in~\eqref{modexp}, which can be evaluated using the stationary phase approximation, 
\begin{equation}
\begin{split}\label{largermodeexp}
A_{z} (u,z,\bz) &= - \frac{i e }{4 \sqrt{2} \pi^{2} }     \int_{0}^{\infty} d \omega \, \left(a_{\CF, +}(\omega, z, \b z)    e^{- i \omega u}  -  a^{\dagger}_{\CF, -}(\omega, z, \b z)    e^{i\omega u} \right)~, \\
\lambda_{(+)} (u,z,\bz) &= - \frac{i e}{ 4\pi^2 } \int_0^\infty d\omega \, \left( 2 \omega \right)^{\frac{1}{2}} \, \left(  a_{\b \Lambda, +}(\omega, z, \b z)     e^{- i \omega u  } - a_{\Lambda, -}^\dagger(\omega, z, \b z)    e^{ i \omega u }    \right) ~, \\
\phi(u,z,\bz) &= - \frac{i}{4\pi^2} \int_0^\infty d\omega \, \left( a_{\b \Phi} (\omega, z, \b z)   e^{ - i \omega u  } - {a}_{\Phi}^{\dagger}(\omega, z, \b z)    e^{ i \omega u  } \right)  ,  \\
\psi (u,z,\bz) &=  - \frac{i}{ 4\pi^2 } \int_0^\infty d\omega \, \left( 2 \omega \right)^{\frac{1}{2}} \,  \left(  a_{\b \Psi, +} (\omega, z, \b z) e^{- i \omega u  } - a_{\Psi, -}^{\dagger} (\omega, z, \b z) e^{ i \omega u }    \right)~.
\end{split}
\end{equation}
Here the momentum~$p^a(\omega, z, \b z)$ of all creation and annihilation operators is parametrized in terms of~$\omega, z, \b z$ as in~\eqref{momparameter}, and we have chosen a gauge such that the photon polarization vectors in~\eqref{modexp} take the following form in flat Minkowski coordinates (see for instance~\cite{hmps}),
\begin{equation}
\ep^{(+) a} (p)  = -{1 \over \sqrt 2} \xi^{(-)\alpha} \,  \sigma^a_{\alpha\betadot} \, \b \xi^{(+)\betadot}(\b z) = {1 \over \sqrt 2} \left(\b z, 1, -i, - \b z\right)~, \qquad \ep^{(-) a}(p) = \left(\ep^{(+)a}(p)\right)^*~.
\end{equation}
We can invert~\eqref{largermodeexp} to express the creation operators in terms of the boundary fields,
\begin{align}
\label{caoperators}
a_{\CF, +}^\dagger    &= - \frac{2\sqrt{2} \pi }{ e \omega } \int du \, e^{-i\omega u} \p_u A_{\b z}~, &  a^\dagger_{\CF, -}   & = - \frac{2\sqrt{2} \pi }{ e \omega } \int du  \, e^{- i\omega u} \, \p_u A_{z}~,    \cr
a^\dagger_{\b \Lambda, +}    &= -\frac{\sqrt{2} \pi i }{ e \sqrt{\omega} } \int du \,  e^{-i \omega u}   \, \b \lambda_{(+)}  ~, &  a_{\Lambda, -}^\dagger   & = - \frac{\sqrt{2} \pi i }{ e \sqrt{\omega} } \int du \, e^{-i \omega u}  \, \lambda_{(+)}~, \cr
a^\dagger_{\b \Phi}   &= - \frac{2\pi}{\omega}   \int du \, e^{-i\omega u} \, \p_u \b \phi  ~, &  a_{\Phi}^\dagger  & = - \frac{2\pi}{\omega}   \int du \, e^{-i\omega u} \, \p_u \phi~,   \cr
a^\dagger_{\b \Psi, +}    &= -\frac{\sqrt{2} \pi i }{  \sqrt{\omega} } \int du \, e^{- i \omega u}  \, \b \psi~, &  a_{\Psi, -}^\dagger   & = - \frac{\sqrt{2} \pi i }{  \sqrt{\omega} } \int du \, e^{-i \omega u}  \, \psi~.   
\end{align}
The corresponding annihilation operators are obtained by Hermitian conjugation.

\bibliography{softphotino-bib}{}

\providecommand{\href}[2]{#2}\begingroup\raggedright\begin{thebibliography}{10}

\bibitem{Weinberg:1995mt}
S.~Weinberg, {\em {The Quantum theory of fields. Vol. 1: Foundations}}.
\newblock Cambridge University Press,
2005.
\newblock

\bibitem{Strominger:2015bla}
A.~Strominger, ``{Magnetic Corrections to the Soft Photon Theorem},''
\href{http://arxiv.org/abs/1509.00543}{{\ttfamily arXiv:1509.00543 [hep-th]}}.

\bibitem{Low:1954kd}
F.~E. Low, ``{Scattering of light of very low frequency by systems of spin
  1/2},''
\href{http://dx.doi.org/10.1103/PhysRev.96.1428}{{\em Phys. Rev.} {\bfseries
  96} (1954) 1428--1432}.

\bibitem{GellMann:1954kc}
M.~Gell-Mann and M.~L. Goldberger, ``{Scattering of low-energy photons by
  particles of spin 1/2},''
\href{http://dx.doi.org/10.1103/PhysRev.96.1433}{{\em Phys. Rev.} {\bfseries
  96} (1954) 1433--1438}.

\bibitem{Low:1958sn}
F.~E. Low, ``{Bremsstrahlung of very low-energy quanta in elementary particle
  collisions},''
\href{http://dx.doi.org/10.1103/PhysRev.110.974}{{\em Phys. Rev.} {\bfseries
  110} (1958) 974--977}.

\bibitem{Weinberg:1965nx}
S.~Weinberg, ``{Infrared photons and gravitons},''
\href{http://dx.doi.org/10.1103/PhysRev.140.B516}{{\em Phys.Rev.} {\bfseries
  140} (1965) B516--B524}.

\bibitem{Burnett:1967km}
T.~H. Burnett and N.~M. Kroll, ``{Extension of the low soft photon theorem},''
\href{http://dx.doi.org/10.1103/PhysRevLett.20.86}{{\em Phys. Rev. Lett.}
  {\bfseries 20} (1968) 86}.

\bibitem{Gross:1968in}
D.~J. Gross and R.~Jackiw, ``{Low-Energy Theorem for Graviton Scattering},''
\href{http://dx.doi.org/10.1103/PhysRev.166.1287}{{\em Phys. Rev.} {\bfseries
  166} (1968) 1287--1292}.

\bibitem{Jackiw:1968zza}
R.~Jackiw, ``{Low-Energy Theorems for Massless Bosons: Photons and
  Gravitons},''
\href{http://dx.doi.org/10.1103/PhysRev.168.1623}{{\em Phys. Rev.} {\bfseries
  168} (1968) 1623--1633}.

\bibitem{White:2011yy}
C.~D. White, ``{Factorization Properties of Soft Graviton Amplitudes},''
  \href{http://dx.doi.org/10.1007/JHEP05(2011)060}{{\em JHEP} {\bfseries 05}
  (2011) 060},
\href{http://arxiv.org/abs/1103.2981}{{\ttfamily arXiv:1103.2981 [hep-th]}}.

\bibitem{Cachazo:2014fwa}
F.~Cachazo and A.~Strominger, ``{Evidence for a New Soft Graviton Theorem},''
\href{http://arxiv.org/abs/1404.4091}{{\ttfamily arXiv:1404.4091 [hep-th]}}.

\bibitem{Casali:2014xpa}
E.~Casali, ``{Soft sub-leading divergences in Yang-Mills amplitudes},''
  \href{http://dx.doi.org/10.1007/JHEP08(2014)077}{{\em JHEP} {\bfseries 08}
  (2014) 077},
\href{http://arxiv.org/abs/1404.5551}{{\ttfamily arXiv:1404.5551 [hep-th]}}.

\bibitem{Schwab:2014xua}
B.~U.~W. Schwab and A.~Volovich, ``{Subleading Soft Theorem in Arbitrary
  Dimensions from Scattering Equations},''
  \href{http://dx.doi.org/10.1103/PhysRevLett.113.101601}{{\em Phys. Rev.
  Lett.} {\bfseries 113} no.~10, (2014) 101601},
\href{http://arxiv.org/abs/1404.7749}{{\ttfamily arXiv:1404.7749 [hep-th]}}.

\bibitem{Bern:2014oka}
Z.~Bern, S.~Davies, and J.~Nohle, ``{On Loop Corrections to Subleading Soft
  Behavior of Gluons and Gravitons},''
  \href{http://dx.doi.org/10.1103/PhysRevD.90.085015}{{\em Phys. Rev.}
  {\bfseries D90} no.~8, (2014) 085015},
\href{http://arxiv.org/abs/1405.1015}{{\ttfamily arXiv:1405.1015 [hep-th]}}.

\bibitem{He:2014bga}
S.~He, Y.-t. Huang, and C.~Wen, ``{Loop Corrections to Soft Theorems in Gauge
  Theories and Gravity},''
  \href{http://dx.doi.org/10.1007/JHEP12(2014)115}{{\em JHEP} {\bfseries 12}
  (2014) 115},
\href{http://arxiv.org/abs/1405.1410}{{\ttfamily arXiv:1405.1410 [hep-th]}}.

\bibitem{Larkoski:2014hta}
A.~J. Larkoski, ``{Conformal Invariance of the Subleading Soft Theorem in Gauge
  Theory},'' \href{http://dx.doi.org/10.1103/PhysRevD.90.087701}{{\em Phys.
  Rev.} {\bfseries D90} no.~8, (2014) 087701},
\href{http://arxiv.org/abs/1405.2346}{{\ttfamily arXiv:1405.2346 [hep-th]}}.

\bibitem{Cachazo:2014dia}
F.~Cachazo and E.~Y. Yuan, ``{Are Soft Theorems Renormalized?},''
\href{http://arxiv.org/abs/1405.3413}{{\ttfamily arXiv:1405.3413 [hep-th]}}.

\bibitem{Afkhami-Jeddi:2014fia}
N.~Afkhami-Jeddi, ``{Soft Graviton Theorem in Arbitrary Dimensions},''
\href{http://arxiv.org/abs/1405.3533}{{\ttfamily arXiv:1405.3533 [hep-th]}}.

\bibitem{Adamo:2014yya}
T.~Adamo, E.~Casali, and D.~Skinner, ``{Perturbative gravity at null
  infinity},'' \href{http://dx.doi.org/10.1088/0264-9381/31/22/225008}{{\em
  Class. Quant. Grav.} {\bfseries 31} no.~22, (2014) 225008},
\href{http://arxiv.org/abs/1405.5122}{{\ttfamily arXiv:1405.5122 [hep-th]}}.

\bibitem{Geyer:2014lca}
Y.~Geyer, A.~E. Lipstein, and L.~Mason, ``{Ambitwistor strings at null infinity
  and (subleading) soft limits},''
  \href{http://dx.doi.org/10.1088/0264-9381/32/5/055003}{{\em Class. Quant.
  Grav.} {\bfseries 32} no.~5, (2015) 055003},
\href{http://arxiv.org/abs/1406.1462}{{\ttfamily arXiv:1406.1462 [hep-th]}}.

\bibitem{Schwab:2014fia}
B.~U.~W. Schwab, ``{Subleading Soft Factor for String Disk Amplitudes},''
  \href{http://dx.doi.org/10.1007/JHEP08(2014)062}{{\em JHEP} {\bfseries 08}
  (2014) 062},
\href{http://arxiv.org/abs/1406.4172}{{\ttfamily arXiv:1406.4172 [hep-th]}}.

\bibitem{Bianchi:2014gla}
M.~Bianchi, S.~He, Y.-t. Huang, and C.~Wen, ``{More on Soft Theorems: Trees,
  Loops and Strings},''
  \href{http://dx.doi.org/10.1103/PhysRevD.92.065022}{{\em Phys. Rev.}
  {\bfseries D92} no.~6, (2015) 065022},
\href{http://arxiv.org/abs/1406.5155}{{\ttfamily arXiv:1406.5155 [hep-th]}}.

\bibitem{Broedel:2014fsa}
J.~Broedel, M.~de~Leeuw, J.~Plefka, and M.~Rosso, ``{Constraining subleading
  soft gluon and graviton theorems},''
  \href{http://dx.doi.org/10.1103/PhysRevD.90.065024}{{\em Phys. Rev.}
  {\bfseries D90} no.~6, (2014) 065024},
\href{http://arxiv.org/abs/1406.6574}{{\ttfamily arXiv:1406.6574 [hep-th]}}.

\bibitem{Bern:2014vva}
Z.~Bern, S.~Davies, P.~Di~Vecchia, and J.~Nohle, ``{Low-Energy Behavior of
  Gluons and Gravitons from Gauge Invariance},''
  \href{http://dx.doi.org/10.1103/PhysRevD.90.084035}{{\em Phys. Rev.}
  {\bfseries D90} no.~8, (2014) 084035},
\href{http://arxiv.org/abs/1406.6987}{{\ttfamily arXiv:1406.6987 [hep-th]}}.

\bibitem{White:2014qia}
C.~D. White, ``{Diagrammatic insights into next-to-soft corrections},''
  \href{http://dx.doi.org/10.1016/j.physletb.2014.08.041}{{\em Phys. Lett.}
  {\bfseries B737} (2014) 216--222},
\href{http://arxiv.org/abs/1406.7184}{{\ttfamily arXiv:1406.7184 [hep-th]}}.

\bibitem{Zlotnikov:2014sva}
M.~Zlotnikov, ``{Sub-sub-leading soft-graviton theorem in arbitrary
  dimension},'' \href{http://dx.doi.org/10.1007/JHEP10(2014)148}{{\em JHEP}
  {\bfseries 10} (2014) 148},
\href{http://arxiv.org/abs/1407.5936}{{\ttfamily arXiv:1407.5936 [hep-th]}}.

\bibitem{Kalousios:2014uva}
C.~Kalousios and F.~Rojas, ``{Next to subleading soft-graviton theorem in
  arbitrary dimensions},''
  \href{http://dx.doi.org/10.1007/JHEP01(2015)107}{{\em JHEP} {\bfseries 01}
  (2015) 107},
\href{http://arxiv.org/abs/1407.5982}{{\ttfamily arXiv:1407.5982 [hep-th]}}.

\bibitem{Du:2014eca}
Y.-J. Du, B.~Feng, C.-H. Fu, and Y.~Wang, ``{Note on Soft Graviton theorem by
  KLT Relation},'' \href{http://dx.doi.org/10.1007/JHEP11(2014)090}{{\em JHEP}
  {\bfseries 11} (2014) 090},
\href{http://arxiv.org/abs/1408.4179}{{\ttfamily arXiv:1408.4179 [hep-th]}}.

\bibitem{Liu:2014vva}
Z.-W. Liu, ``{Soft theorems in maximally supersymmetric theories},''
  \href{http://dx.doi.org/10.1140/epjc/s10052-015-3304-1}{{\em Eur. Phys. J.}
  {\bfseries C75} no.~3, (2015) 105},
\href{http://arxiv.org/abs/1410.1616}{{\ttfamily arXiv:1410.1616 [hep-th]}}.

\bibitem{Rao:2014zaa}
J.~Rao, ``{Soft theorem of $ \mathcal{N} $ = 4 SYM in Grassmannian
  formulation},'' \href{http://dx.doi.org/10.1007/JHEP02(2015)087}{{\em JHEP}
  {\bfseries 02} (2015) 087},
\href{http://arxiv.org/abs/1410.5047}{{\ttfamily arXiv:1410.5047 [hep-th]}}.

\bibitem{Bonocore:2014wua}
D.~Bonocore, E.~Laenen, L.~Magnea, L.~Vernazza, and C.~D. White, ``{The method
  of regions and next-to-soft corrections in Drell--Yan production},''
  \href{http://dx.doi.org/10.1016/j.physletb.2015.02.008}{{\em Phys. Lett.}
  {\bfseries B742} (2015) 375--382},
\href{http://arxiv.org/abs/1410.6406}{{\ttfamily arXiv:1410.6406 [hep-ph]}}.

\bibitem{Luo:2014wea}
H.~Luo, P.~Mastrolia, and W.~J. Torres~Bobadilla, ``{Subleading soft behavior
  of QCD amplitudes},''
  \href{http://dx.doi.org/10.1103/PhysRevD.91.065018}{{\em Phys. Rev.}
  {\bfseries D91} no.~6, (2015) 065018},
\href{http://arxiv.org/abs/1411.1669}{{\ttfamily arXiv:1411.1669 [hep-th]}}.

\bibitem{Broedel:2014bza}
J.~Broedel, M.~de~Leeuw, J.~Plefka, and M.~Rosso, ``{Local contributions to
  factorized soft graviton theorems at loop level},''
  \href{http://dx.doi.org/10.1016/j.physletb.2015.05.018}{{\em Phys. Lett.}
  {\bfseries B746} (2015) 293--299},
\href{http://arxiv.org/abs/1411.2230}{{\ttfamily arXiv:1411.2230 [hep-th]}}.

\bibitem{Schwab:2014sla}
B.~U.~W. Schwab, ``{A Note on Soft Factors for Closed String Scattering},''
  \href{http://dx.doi.org/10.1007/JHEP03(2015)140}{{\em JHEP} {\bfseries 03}
  (2015) 140},
\href{http://arxiv.org/abs/1411.6661}{{\ttfamily arXiv:1411.6661 [hep-th]}}.

\bibitem{Chen:2014xoa}
W.-M. Chen, Y.-t. Huang, and C.~Wen, ``{New Fermionic Soft Theorems for
  Supergravity Amplitudes},''
  \href{http://dx.doi.org/10.1103/PhysRevLett.115.021603}{{\em Phys. Rev.
  Lett.} {\bfseries 115} no.~2, (2015) 021603},
\href{http://arxiv.org/abs/1412.1809}{{\ttfamily arXiv:1412.1809 [hep-th]}}.

\bibitem{Chen:2014cuc}
W.-M. Chen, Y.-t. Huang, and C.~Wen, ``{From U(1) to E8: soft theorems in
  supergravity amplitudes},''
  \href{http://dx.doi.org/10.1007/JHEP03(2015)150}{{\em JHEP} {\bfseries 03}
  (2015) 150},
\href{http://arxiv.org/abs/1412.1811}{{\ttfamily arXiv:1412.1811 [hep-th]}}.

\bibitem{Larkoski:2014bxa}
A.~J. Larkoski, D.~Neill, and I.~W. Stewart, ``{Soft Theorems from Effective
  Field Theory},'' \href{http://dx.doi.org/10.1007/JHEP06(2015)077}{{\em JHEP}
  {\bfseries 06} (2015) 077},
\href{http://arxiv.org/abs/1412.3108}{{\ttfamily arXiv:1412.3108 [hep-th]}}.

\bibitem{Vera:2014tda}
A.~Sabio~Vera and M.~A. Vazquez-Mozo, ``{The Double Copy Structure of Soft
  Gravitons},'' \href{http://dx.doi.org/10.1007/JHEP03(2015)070}{{\em JHEP}
  {\bfseries 03} (2015) 070},
\href{http://arxiv.org/abs/1412.3699}{{\ttfamily arXiv:1412.3699 [hep-th]}}.

\bibitem{DiVecchia:2015oba}
P.~Di~Vecchia, R.~Marotta, and M.~Mojaza, ``{Soft theorem for the graviton,
  dilaton and the Kalb-Ramond field in the bosonic string},''
  \href{http://dx.doi.org/10.1007/JHEP05(2015)137}{{\em JHEP} {\bfseries 05}
  (2015) 137},
\href{http://arxiv.org/abs/1502.05258}{{\ttfamily arXiv:1502.05258 [hep-th]}}.

\bibitem{Cachazo:2015ksa}
F.~Cachazo, S.~He, and E.~Y. Yuan, ``{New Double Soft Emission Theorems},''
  \href{http://dx.doi.org/10.1103/PhysRevD.92.065030}{{\em Phys. Rev.}
  {\bfseries D92} no.~6, (2015) 065030},
\href{http://arxiv.org/abs/1503.04816}{{\ttfamily arXiv:1503.04816 [hep-th]}}.

\bibitem{Lipstein:2015rxa}
A.~E. Lipstein, ``{Soft Theorems from Conformal Field Theory},''
  \href{http://dx.doi.org/10.1007/JHEP06(2015)166}{{\em JHEP} {\bfseries 06}
  (2015) 166},
\href{http://arxiv.org/abs/1504.01364}{{\ttfamily arXiv:1504.01364 [hep-th]}}.

\bibitem{Adamo:2015fwa}
T.~Adamo and E.~Casali, ``{Perturbative gauge theory at null infinity},''
  \href{http://dx.doi.org/10.1103/PhysRevD.91.125022}{{\em Phys. Rev.}
  {\bfseries D91} no.~12, (2015) 125022},
\href{http://arxiv.org/abs/1504.02304}{{\ttfamily arXiv:1504.02304 [hep-th]}}.

\bibitem{Klose:2015xoa}
T.~Klose, T.~McLoughlin, D.~Nandan, J.~Plefka, and G.~Travaglini,
  ``{Double-Soft Limits of Gluons and Gravitons},''
  \href{http://dx.doi.org/10.1007/JHEP07(2015)135}{{\em JHEP} {\bfseries 07}
  (2015) 135},
\href{http://arxiv.org/abs/1504.05558}{{\ttfamily arXiv:1504.05558 [hep-th]}}.

\bibitem{Volovich:2015yoa}
A.~Volovich, C.~Wen, and M.~Zlotnikov, ``{Double Soft Theorems in Gauge and
  String Theories},'' \href{http://dx.doi.org/10.1007/JHEP07(2015)095}{{\em
  JHEP} {\bfseries 07} (2015) 095},
\href{http://arxiv.org/abs/1504.05559}{{\ttfamily arXiv:1504.05559 [hep-th]}}.

\bibitem{Bianchi:2015yta}
M.~Bianchi and A.~L. Guerrieri, ``{On the soft limit of open string disk
  amplitudes with massive states},''
  \href{http://dx.doi.org/10.1007/JHEP09(2015)164}{{\em JHEP} {\bfseries 09}
  (2015) 164},
\href{http://arxiv.org/abs/1505.05854}{{\ttfamily arXiv:1505.05854 [hep-th]}}.

\bibitem{Bork:2015fla}
L.~V. Bork and A.~I. Onishchenko, ``{On Soft Theorems And Form Factors In N=4
  SYM Theory},''
\href{http://arxiv.org/abs/1506.07551}{{\ttfamily arXiv:1506.07551 [hep-th]}}.

\bibitem{DiVecchia:2015bfa}
P.~Di~Vecchia, R.~Marotta, and M.~Mojaza, ``{Double-soft behavior for scalars
  and gluons from string theory},''
\href{http://arxiv.org/abs/1507.00938}{{\ttfamily arXiv:1507.00938 [hep-th]}}.

\bibitem{Guerrieri:2015eea}
A.~L. Guerrieri, ``{Soft behavior of string amplitudes with external massive
  states},'' in {\em {27th Conference on High Energy Physics (IFAE 2015) Rome,
  Italy, April 8-10, 2015}}.
\newblock 2015.
\newblock
\href{http://arxiv.org/abs/1507.08829}{{\ttfamily arXiv:1507.08829 [hep-th]}}.
\newblock

\bibitem{Alston:2015gea}
S.~D. Alston, D.~C. Dunbar, and W.~B. Perkins, ``{$n$-point amplitudes with a
  single negative-helicity graviton},''
  \href{http://dx.doi.org/10.1103/PhysRevD.92.065024}{{\em Phys. Rev.}
  {\bfseries D92} no.~6, (2015) 065024},
\href{http://arxiv.org/abs/1507.08882}{{\ttfamily arXiv:1507.08882 [hep-th]}}.

\bibitem{Chin:2015qza}
S.~Chin, S.~Lee, and Y.~Yun, ``{ABJM Amplitudes in U-gauge and a Soft
  Theorem},''
\href{http://arxiv.org/abs/1508.07975}{{\ttfamily arXiv:1508.07975 [hep-th]}}.

\bibitem{DiVecchia:2015srk}
P.~Di~Vecchia, R.~Marotta, and M.~Mojaza, ``{Soft Theorems from String
  Theory},''
\newblock 2015.
\newblock
\href{http://arxiv.org/abs/1511.04921}{{\ttfamily arXiv:1511.04921 [hep-th]}}.
\newblock

\bibitem{Brandhuber:2015vhm}
A.~Brandhuber, E.~Hughes, B.~Spence, and G.~Travaglini, ``{One-Loop Soft
  Theorems via Dual Superconformal Symmetry},''
\href{http://arxiv.org/abs/1511.06716}{{\ttfamily arXiv:1511.06716 [hep-th]}}.

\bibitem{as}
A.~Strominger, ``{Asymptotic Symmetries of Yang-Mills Theory},''
  \href{http://dx.doi.org/10.1007/JHEP07(2014)151}{{\em JHEP} {\bfseries 07}
  (2014) 151},
\href{http://arxiv.org/abs/1308.0589}{{\ttfamily arXiv:1308.0589 [hep-th]}}.

\bibitem{as1}
A.~Strominger, ``{On BMS Invariance of Gravitational Scattering},''
  \href{http://dx.doi.org/10.1007/JHEP07(2014)152}{{\em JHEP} {\bfseries 07}
  (2014) 152},
\href{http://arxiv.org/abs/1312.2229}{{\ttfamily arXiv:1312.2229 [hep-th]}}.

\bibitem{He:2014laa}
T.~He, V.~Lysov, P.~Mitra, and A.~Strominger, ``{BMS supertranslations and
  Weinberg's soft graviton theorem},''
  \href{http://dx.doi.org/10.1007/JHEP05(2015)151}{{\em JHEP} {\bfseries 05}
  (2015) 151},
\href{http://arxiv.org/abs/1401.7026}{{\ttfamily arXiv:1401.7026 [hep-th]}}.

\bibitem{Kapec:2014opa}
D.~Kapec, V.~Lysov, S.~Pasterski, and A.~Strominger, ``{Semiclassical Virasoro
  symmetry of the quantum gravity $ \mathcal{S}$-matrix},''
  \href{http://dx.doi.org/10.1007/JHEP08(2014)058}{{\em JHEP} {\bfseries 08}
  (2014) 058},
\href{http://arxiv.org/abs/1406.3312}{{\ttfamily arXiv:1406.3312 [hep-th]}}.

\bibitem{hmps}
T.~He, P.~Mitra, A.~P. Porfyriadis, and A.~Strominger, ``{New Symmetries of
  Massless QED},'' \href{http://dx.doi.org/10.1007/JHEP10(2014)112}{{\em JHEP}
  {\bfseries 10} (2014) 112},
\href{http://arxiv.org/abs/1407.3789}{{\ttfamily arXiv:1407.3789 [hep-th]}}.

\bibitem{Lysov:2014csa}
V.~Lysov, S.~Pasterski, and A.~Strominger, ``{Low's Subleading Soft Theorem as
  a Symmetry of QED},''
  \href{http://dx.doi.org/10.1103/PhysRevLett.113.111601}{{\em Phys. Rev.
  Lett.} {\bfseries 113} no.~11, (2014) 111601},
\href{http://arxiv.org/abs/1407.3814}{{\ttfamily arXiv:1407.3814 [hep-th]}}.

\bibitem{Campiglia:2014yka}
M.~Campiglia and A.~Laddha, ``{Asymptotic symmetries and subleading soft
  graviton theorem},'' \href{http://dx.doi.org/10.1103/PhysRevD.90.124028}{{\em
  Phys. Rev.} {\bfseries D90} no.~12, (2014) 124028},
\href{http://arxiv.org/abs/1408.2228}{{\ttfamily arXiv:1408.2228 [hep-th]}}.

\bibitem{Kapec:2014zla}
D.~Kapec, V.~Lysov, and A.~Strominger, ``{Asymptotic Symmetries of Massless QED
  in Even Dimensions},''
\href{http://arxiv.org/abs/1412.2763}{{\ttfamily arXiv:1412.2763 [hep-th]}}.

\bibitem{Mohd:2014oja}
A.~Mohd, ``{A note on asymptotic symmetries and soft-photon theorem},''
  \href{http://dx.doi.org/10.1007/JHEP02(2015)060}{{\em JHEP} {\bfseries 02}
  (2015) 060},
\href{http://arxiv.org/abs/1412.5365}{{\ttfamily arXiv:1412.5365 [hep-th]}}.

\bibitem{Campiglia:2015yka}
M.~Campiglia and A.~Laddha, ``{New symmetries for the Gravitational
  S-matrix},'' \href{http://dx.doi.org/10.1007/JHEP04(2015)076}{{\em JHEP}
  {\bfseries 04} (2015) 076},
\href{http://arxiv.org/abs/1502.02318}{{\ttfamily arXiv:1502.02318 [hep-th]}}.

\bibitem{Kapec:2015vwa}
D.~Kapec, V.~Lysov, S.~Pasterski, and A.~Strominger, ``{Higher-Dimensional
  Supertranslations and Weinberg's Soft Graviton Theorem},''
\href{http://arxiv.org/abs/1502.07644}{{\ttfamily arXiv:1502.07644 [gr-qc]}}.

\bibitem{He:2015zea}
T.~He, P.~Mitra, and A.~Strominger, ``{2D Kac-Moody Symmetry of 4D Yang-Mills
  Theory},''
\href{http://arxiv.org/abs/1503.02663}{{\ttfamily arXiv:1503.02663 [hep-th]}}.

\bibitem{Campiglia:2015qka}
M.~Campiglia and A.~Laddha, ``{Asymptotic symmetries of QED and Weinberg?s soft
  photon theorem},'' \href{http://dx.doi.org/10.1007/JHEP07(2015)115}{{\em
  JHEP} {\bfseries 07} (2015) 115},
\href{http://arxiv.org/abs/1505.05346}{{\ttfamily arXiv:1505.05346 [hep-th]}}.

\bibitem{Kapec:2015ena}
D.~Kapec, M.~Pate, and A.~Strominger, ``{New Symmetries of QED},''
\href{http://arxiv.org/abs/1506.02906}{{\ttfamily arXiv:1506.02906 [hep-th]}}.

\bibitem{Avery:2015gxa}
S.~G. Avery and B.~U.~W. Schwab, ``{BMS, String Theory, and Soft Theorems},''
\href{http://arxiv.org/abs/1506.05789}{{\ttfamily arXiv:1506.05789 [hep-th]}}.

\bibitem{Campiglia:2015kxa}
M.~Campiglia and A.~Laddha, ``{Asymptotic symmetries of gravity and soft
  theorems for massive particles},''
\href{http://arxiv.org/abs/1509.01406}{{\ttfamily arXiv:1509.01406 [hep-th]}}.

\bibitem{Avery:2015rga}
S.~G. Avery and B.~U.~W. Schwab, ``{Noether's Second Theorem and Ward
  Identities for Gauge Symmetries},''
\href{http://arxiv.org/abs/1510.07038}{{\ttfamily arXiv:1510.07038 [hep-th]}}.

\bibitem{Sakai:1989nh}
N.~Sakai and Y.~Tanii, ``{Super Wess-Zumino-Witten Models From Super
  Chern-Simons Theories},''
\href{http://dx.doi.org/10.1143/PTP.83.968}{{\em Prog. Theor. Phys.} {\bfseries
  83} (1990) 968--990}.

\bibitem{ly}
V.~Lysov, ``{Asymptotical Supersymmetries from Gravitino Soft Theorem}.'' To
  appear.

\bibitem{avsch}
S.~G. Avery and B.~U.~W. Schwab. In progress.

\bibitem{Wess:1992cp}
J.~Wess and J.~Bagger, {\em {Supersymmetry and supergravity}}.
\newblock
1992.
\newblock

\bibitem{Kapec:2017tkm}
D.~Kapec, M.~Perry, A.-M. Raclariu, and A.~Strominger, ``{Infrared Divergences
  in QED, Revisited},''
  \href{http://dx.doi.org/10.1103/PhysRevD.96.085002}{{\em Phys. Rev. D}
  {\bfseries 96} no.~8, (2017) 085002},
  \href{http://arxiv.org/abs/1705.04311}{{\ttfamily arXiv:1705.04311
  [hep-th]}}.

\bibitem{Hofman:2008ar}
D.~M. Hofman and J.~Maldacena, ``{Conformal collider physics: Energy and charge
  correlations},'' \href{http://dx.doi.org/10.1088/1126-6708/2008/05/012}{{\em
  JHEP} {\bfseries 05} (2008) 012},
\href{http://arxiv.org/abs/0803.1467}{{\ttfamily arXiv:0803.1467 [hep-th]}}.

\bibitem{He:2020ifr}
T.~He and P.~Mitra, ``{Covariant Phase Space and Soft Factorization in
  Non-abelian Gauge Theories},''
  \href{http://arxiv.org/abs/2009.14334}{{\ttfamily arXiv:2009.14334
  [hep-th]}}.

\end{thebibliography}\endgroup
\bibliographystyle{plain}

\end{document}